\documentclass[letterpaper,twocolumn,10pt]{article}
\usepackage{usenix-2020-09}

\usepackage{tikz}
\usepackage{amsmath}
\usepackage{microtype}
\usepackage{graphicx}
\usepackage{subcaption}
\usepackage{booktabs} %
\usepackage{array}    %
\usepackage{multirow}
\usepackage{float}

\usepackage{hyperref}
\usepackage{enumitem}

\newcommand{\smartparagraph}[1]{\noindent{\bf #1}\ }
\newif\ifhidecomments
\hidecommentsfalse  %
\ifhidecomments
  \newcommand{\satya}[1]{}
  \newcommand{\ag}[1]{}
\else
  \newcommand{\satya}[1]{\textcolor{blue}{Satya: #1}}
  \newcommand{\ag}[1]{\textcolor{magenta}{AG: #1}}
  
\fi

\newcommand{\sys}{\textsc{NetBurst}~}

\begin{document}

\date{}

\title{\sys: Event-Centric Forecasting \\ of Bursty, Intermittent Time Series}

\author{%
\begin{minipage}[t]{0.32\textwidth}\centering
\bfseries
Satyandra Guthula\textsuperscript{1}\\[1ex]
Kesheng Wu\textsuperscript{3}
\end{minipage}\hfill
\begin{minipage}[t]{0.32\textwidth}\centering
\bfseries
Jaber Daneshamooz\textsuperscript{1}\\[1ex]
Walter Willinger\textsuperscript{4}
\end{minipage}\hfill
\begin{minipage}[t]{0.32\textwidth}\centering
\bfseries
Charles Fleming\textsuperscript{2}\\[1ex]
Arpit Gupta\textsuperscript{1}
\end{minipage}
\\[10pt]
\normalfont
\textsuperscript{1}University of California, Santa Barbara\\[3pt]
\begingroup
\setlength{\tabcolsep}{0pt}%
\begin{tabular*}{\textwidth}{@{\extracolsep{\fill}} l l l}
\textsuperscript{2}Cisco Research &
\textsuperscript{3}Lawrence Berkeley National Laboratory &
\textsuperscript{4}Northwestern University
\end{tabular*}
\endgroup
}

\maketitle

\begin{abstract}
Network operators monitor their infrastructure by collecting telemetry data such as packet counts, byte rates, or flow volumes, yet answering the questions that effective operations demand---forecasting future load, diagnosing and characterizing anomalies, and searching for and retrieving historical precedents---requires more than raw measurements.
Bridging this gap calls for learned representations: compact per-entity summaries that capture temporal dynamics from each entity's univariate time series.
Time-series foundation models are the natural starting point, but they are designed for dense, periodic benchmark datasets---the \emph{mild} statistical regime. However,
network telemetry data inhabits the \emph{wild} regime: operationally relevant events are rare, separated by variable-length stretches of low or no activity (``ebbs''), with intermittent bursts of heavy-tailed extremes (``tides'').
We present \sys, an event-centric pipeline that collapses ebbs, separates each time series into a stream of burst timings and a stream of burst magnitudes, and learns a single representation serving all three operational tasks.
Compared to the strongest competitors among eight baselines---including Amazon's Chronos-2 and Datadog's Toto---and across nine production telemetry configurations, \sys reduces median forecasting error by $1.3$--$116\times$ on wild-regime data with a $1.0$--$7.5\times$ better match to the true burst distribution, and matches baselines on mild-regime benchmarks.
For characterizing anomalies, \sys produces balanced, well-spread clusters that are $16\times$ more describable in operator-familiar terms under a novel interpretability score, and cluster-filtered search delivers $7.5\times$ faster end-to-end retrieval.
\end{abstract}
\section{Introduction}
\label{sec:introduction}

\smartparagraph{Network telemetry data.}
Network operators monitor their infrastructure by collecting telemetry data---packet counts, byte rates, latency samples, device counters---at granularities from individual services and IP addresses to entire subnets, and at temporal resolutions from milliseconds to hours~\cite{guo2015pingmesh, zhou2020pcat, miao2022optel}.
This data underpins core operational tasks.
A backbone operator scheduling a large data transfer uses it to forecast whether a cross-country link will approach saturation (\emph{forecasting}).
A campus engineer investigating a latency spike relies on it to determine whether it matches a known pattern---a congestion event, a misconfiguration, a DDoS attack---or represents something the network has never seen (\emph{diagnosis}).
And it is critical for an incident responder diagnosing correlated anomalies who wants to retrieve the last instance when this pattern appeared across the same set of nodes (\emph{semantic search}).

However, raw telemetry data resists direct use for three reasons.
It is \emph{voluminous}: a single campus gateway can produce tens of millions of packet-count and byte-rate measurements per hour across thousands of monitored endpoints~\cite{beltiukov2023pinot, cho_traffic_2000}.
It is \emph{sparse}: for much of the time, most endpoints are idle, and the rare bursts of activity that matter are buried in long stretches of low activity (noise) or no activity (zeros)~\cite{wierman2023heavytails, leland2002self}.
And it is \emph{semantically opaque}: a raw byte-count time series leaves the operator guessing whether a spike is a congestion event, an attack-related anomaly, or routine traffic---and deployed anomaly-detection tools that flag deviations without characterizing their type~\cite{mueller2025} do not close this gap.
This paper bridges the gap between what operators have---raw telemetry data---and what effective network operations demand: answers to questions that concern tasks such as forecasting, diagnosis, and semantic search.

\smartparagraph{Representation learning for network telemetry.}
Bridging this gap requires learning the spatial and temporal relationships latent in telemetry data: how traffic for each monitored entity---an IP address, a service, a subnet---evolves over time, and how patterns correlate across entities.
Rather than building a separate model for each operational task, operators benefit from a single compact summary of each entity's behavior---a \emph{learned representation}~\cite{bengio2013representation}: a fixed-size numerical vector that captures an entity's temporal dynamics while concentrating on the rare events that carry operational meaning.
Downstream tools consume such a vector for forecasting, diagnosis, and semantic search alike.
Learning representations jointly across all entities captures spatial correlations but couples them into a single model whose cost grows with the number of monitored endpoints~\cite{cao2020stemgnn, transformer_hawkes}.
A modular alternative first learns a temporal representation of each entity's univariate time series independently~\cite{nie2023patchtst}, then composes spatial relationships on top as needed~\cite{liu2024itransformer}.
This paper focuses on the first layer: learning temporal representations from univariate time series that include per-entity packet counts, byte rates, or flow volumes.
Recent network foundation models learn such representations from packet-level traces~\cite{guthula_netfound_2025, lin_et-bert_2022, wang_lens_2024, zhao_debunking_2025}, capturing fine-grained protocol structure but operating at a granularity too detailed for most operational tasks that reason about aggregate counts and rates at the entity level.
Time-series foundation models (TS-FMs)~\cite{ansari2024chronos, das2024timesfm, lagllama} operate at this aggregate level---learning temporal representations that capture dependencies hand-crafted statistical features miss and that LLMs cannot economically extract across each of the thousands of monitored entities simultaneously---and are the natural starting point for the problem this paper addresses.

\smartparagraph{Problems with time-series foundation models.}
However, these models face two problems when applied to network telemetry data: the data they expect and the objective they optimize.
Existing TS-FMs are designed for benchmark datasets---exchange rates, electricity loads, taxi demands~\cite{zhou2021informer, electricityloaddiagrams20112014_321}---where observations are dense, periodic, and moderately variable. Mandelbrot~\cite{mandelbrot1997} called this the \emph{mild} statistical regime.
Network telemetry inhabits what he termed the \emph{wild} regime~\cite{wierman2023heavytails}: the events operators care about---congestion bursts, anomalous spikes, capacity shifts---are rare, separated by long idle stretches, and their sizes can be orders of magnitude larger than the mean.
These models process a fixed input window---512 data points in Chronos~\cite{ansari2024chronos}, for example. When a sparse telemetry time series has fewer than ten operationally meaningful events in those 512 slots, the model sees mostly silence. Extending the window to capture more events is quadratically expensive~\cite{vaswani_attention_2017}; the model spends most of its capacity learning to predict silence rather than the rare events that matter in practice (\S\ref{sec:motivation}).
Two design choices make this worse: the models predict burst timing and burst magnitude with the same set of parameters---forcing one model to fit two problems whose errors compound---and they give common small values and rare extreme values the same resolution, washing out the large events operators care about most~\cite{ansari2024chronos}.

Even when these models forecast adequately, accurate prediction does not guarantee useful representations: a model can predict well by memorizing recent trends yet organize its internal summaries so that behaviorally similar time series are scattered and dissimilar ones land nearby in the representation space~\cite{beltiukov2025demystifying}.
A common property of existing models is that they concentrate their representations into narrow cones---most vectors point in nearly the same direction~\cite{beltiukov2025demystifying}---so that distance between two time series' summaries no longer reflects behavioral similarity, undermining both diagnosis and search.
We confirm this property in \S\ref{subsec:repr_quality}: baselines that forecast competitively still collapse to a single effective dimension.

\smartparagraph{Our approach: event-centric representations.}
Network telemetry data alternates between long idle stretches and sudden bursts of activity---ebbs and tides.
Our approach concentrates the model's input on the tides: bursts that rise above the noise floor, compressing the idle stretches between them (\emph{eventization}).
Each burst carries two operationally distinct signals: \emph{when} it occurs and \emph{how large} it is.
Rather than forcing both into a single model, we decompose each time series into an inter-burst gap stream and a burst intensity stream.
To preserve the tail events that uniform binning erases~\cite{ansari2024chronos}, we discretize each stream so that rare extreme values receive the same resolution as common ones (\emph{quantile tokenization}).
A shared encoder with a dedicated prediction head per stream captures cross-stream dependencies while disentangling timing and magnitude at the output, producing representations (embeddings) that spread evenly across the representation space so that geometric distance between two time series reflects behavioral similarity (\emph{twin-head encoder}).
A characterization pipeline maps each cluster to operator-familiar statistical features (e.g., TSFresh~\cite{tsfresh}) that both align with the embedding and discriminate across clusters.
Together, these components form a single architecture that we fine-tune per granularity to serve forecasting, diagnosis, and semantic search on wild-regime telemetry data where existing models fail, while remaining competitive on mild-regime data.

\begin{table*}[t!]
    \centering
    \small
    \resizebox{\linewidth}{!}{%
    \begin{tabular}{l|rrrrrrrrr|r}
        & \textbf{\sys} & \textbf{Chronos}~\cite{ansari2024chronos} & \textbf{Chronos-2}~\cite{ansari2025chronos2} & \textbf{DeepAR}~\cite{deepar} & \textbf{Lag-Llama}~\cite{lagllama} & \textbf{Toto}~\cite{toto} & \textbf{N-BEATS}~\cite{n-beats} & \textbf{THP}~\cite{transformer_hawkes} & \textbf{S2P2}~\cite{chang2025s2p2} & \textbf{VMR} \\
        \hline
        Exchange Rate (daily) & 0.6980 & 0.0341 & 0.0242 & 0.0833 & 0.0800 & \textbf{0.0183} & 0.0411 & 0.1168 & 1.1477 & 0.02\\
        ETT-M2 (hourly) & 0.5413 & \textbf{0.3717} & 0.3937 & 0.5538 & 0.9276 & 0.4947 & 0.3874 & 1.2519 & 1.0058 & 3.6\\
        Taxi (30m) & 0.5789 & 0.5747 & \textbf{0.5143} & 0.6736 & 1.4770 & 1.3646 & 0.6645 & 0.5907 & 4.3430 & 4.7\\
        Weather (daily) & 0.9345 & 0.2590 & \textbf{0.2375} & 0.2899 & 0.2742 & 0.2991 & 0.6971 & 0.6238 & 1.0017 & 29\\
        Electricity (15m) & 0.1116 & 0.0895 & \textbf{0.0700} & 0.1455 & 0.1779 & 0.3362 & 0.9999 & 3.5708 & 0.9999 & 35\\
        \hline
        Campus-Pkts-1 (service, 100ms) & \textbf{0.0083} & 1.0000 & 0.9971 & 1.0002 & 1.0000 & 0.9710 & 0.9882 & 1.2836 & 1.0000 & 102\\
        Campus-Pkts-1 (IP, 1s) & \textbf{0.5652} & 0.9979 & 0.9768 & 1.0002 & 1.0000 & 1.3380 & 2.4770 & 38.6351 & 1.0000 & 1.2k\\
        Campus-Pkts-1 (/24, 1s) & \textbf{0.6004} & 0.9926 & 0.9706 & 1.0000 & 1.0000 & 1.5109 & 0.9653 & 22.3202 & 1.0000 & 1.3k\\
        MAWI (service, 100ms) & \textbf{0.0093} & 1.0000 & 0.9987 & 0.9995 & 1.0000 & 0.9730 & 0.8390 & 5.5073 & 1.0000 & 68\\
        MAWI (IP, 1s) & \textbf{0.4568} & 0.9778 & 0.9913 & 0.9999 & 1.0000 & 1.1687 & 0.8900 & 36.6185 & 1.0000 & 307\\
        MAWI (/24, 1s) & \textbf{0.5214} & 0.9285 & 0.9288 & 1.0000 & 1.0000 & 1.0589 & 0.8227 & 59.0166 & 1.0000 & 1.1k\\
        RnE-NetFlow (IP, 60s) & \textbf{0.5855} & 0.8794 & 0.7420 & 0.9490 & 0.8597 & 1.6287 & 1.0493 & 0.9487 & 1.0000 & 59.0M\\
        \hline
              \hline
        \textbf{Wild/mild ratio} & 0.1$\times$ & 25.8$\times$ & 30.7$\times$ & 11.4$\times$ & 10.7$\times$ & 53.1$\times$ & 20.0$\times$ & 8.1$\times$ & 1.0$\times$ & \\
    \end{tabular}%
    }
    \caption{Median MAPE (Mean Absolute Percentage Error, where $1.0 = 100\%$ error; values above~$1$ indicate predictions that differ from the true value by more than $100\%$) on mild-regime benchmarks (top) vs.\ wild-regime telemetry (bottom). Each cell reports the median MAPE across all entities in a dataset; the rightmost column reports median VMR (variance-to-mean ratio), an indicator of statistical regime membership (\S\ref{sec:motivation}). The bottom row reports the most conservative degradation ratio for each model: its lowest wild-regime MAPE divided by its lowest mild-regime MAPE.}
    \label{tab:temporal_baselines}
    \vspace{-1.0em}
\end{table*}

\smartparagraph{Contributions.} The paper makes the following contributions:
\begin{enumerate}[leftmargin=1.2em, itemsep=2pt, topsep=2pt, parsep=0pt]
    \item \textbf{Statistical regime characterization.} We show that network telemetry time series inhabit a qualitatively different statistical regime from standard benchmark time series~\cite{mandelbrot1997}---sparse, event-driven, and heavy-tailed---and that this regime shift breaks existing baselines: even in the best case, forecasting errors on wild-regime data are up to $53\times$ larger than on mild-regime benchmarks (Table~\ref{tab:temporal_baselines}, Section~\ref{sec:motivation}).
    \item \textbf{Event-centric pipeline.} We introduce \sys, a novel event-centric pipeline---eventization, quantile tokenization, and a twin-head encoder---that focuses on burst events, disentangles burst timing from magnitude, and produces well-spread embeddings that cluster into operator-readable behavioral clusters. To quantify how describable each cluster is, we introduce an interpretability score that measures whether a cluster's defining statistical features are both consistent with its embedding and distinctive from other clusters (Section~\ref{sec:design}).
    \item \textbf{Forecasting, diagnosis, and search on production telemetry data.} Compared to the strongest competitors among eight baselines and across nine telemetry configurations, \sys achieves $1.3$--$116\times$ lower median forecasting error on wild-regime data with $1.0$--$7.5\times$ better match to the true burst distribution, matches baselines on mild-regime data, and produces clusters with $16\times$ higher median interpretability score. Cluster-filtered search delivers $7.5\times$ faster end-to-end retrieval (Section~\ref{sec:evaluation}).
\end{enumerate}
\noindent On acceptance, we will open-source \sys and release all anonymized data and code for reproducibility. The details of the code and artifacts are in Appendix \ref{sec:availability}.

\vspace{-1em}

\section{Background and Motivation}
\label{sec:motivation}

Applying time-series foundation models to network telemetry data exposes two problems: the data breaks the models' underlying assumptions, and the models' training objective targets forecasting alone, ignoring the geometric structure that diagnosis and search require.

\smartparagraph{Problem 1: A data mismatch.}
Whether an operator monitors individual applications or entire backbone links, the same overall pattern holds: most of the time, most entities---individual services, IP addresses, subnets, or links that an operator monitors independently---exhibit routine behavior that warrants no action.
A service-level packet trace at 100\,ms granularity makes this evident---most time bins are empty, and the rare bursts stand out against silence.
A subnet-level byte-rate trace at 1\,s or a backbone NetFlow record at 60\,s looks different on the surface: every bin carries some traffic, and the time series appears dense and continuous, yet with discernible ``ebbs" and ``tides".
But the operator's question is the same: where are and how large are the departures that warrant attention---a congestion burst, a latency spike, a traffic shift that precedes a link failure?
The same pattern recurs across modalities: latency probes~\cite{guo2015pingmesh} fluctuate continuously, optical power levels~\cite{miao2022optel} drift gradually, device counters~\cite{zhou2020pcat} tick steadily---yet in each case, the operationally relevant signal lives in the rare departures from the signal's baseline.
Whether the trace is already sparse or an operator-determined threshold separates the routine noise floor from the events that matter, the end result is the same: a sparse sequence of operationally relevant events separated by stretches of silence or irrelevance.

This event-driven structure sets network telemetry data apart from the data existing models were designed for.
Mandelbrot~\cite{mandelbrot1997} distinguished two qualitatively different statistical regimes: \emph{mild}, where fluctuations are moderate and distributions are well-approximated by Gaussians~\cite{deepar}, and \emph{wild}, where rare extreme events dominate the magnitude and variability of the process~\cite{wierman2023heavytails}.
Standard benchmarks---exchange rates, electricity loads, taxi demand~\cite{zhou2021informer, electricityloaddiagrams20112014_321, ny_taxi, exchange_rate}---inhabit the mild regime: dense, approximately stationary, with predictable cyclical structure.
The sparse, bursty event structure of network telemetry data places it in the wild regime: event magnitudes follow heavy-tailed distributions~\cite{Nair2013TheFO}, and variance estimates converge slowly or not at all.

\begin{figure}[h]
    \centering
    \includegraphics[width=.9\linewidth]{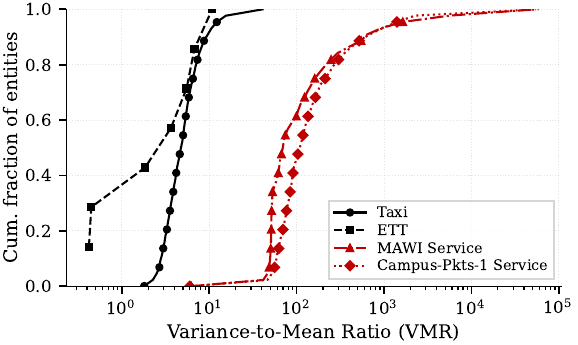}
    \caption{VMR distributions for standard benchmarks (low variability, narrow range) and network telemetry datasets (high variability, spanning $10^{2}$--$10^{4}$).}
    \label{fig:fano_ccdf}
    \label{fig:telemetry_stats}
    \vspace{-1.0em}
\end{figure}

The variance-to-mean ratio (VMR) serves as an empirical indicator of this regime membership: VMR near~$1$ indicates predictable, Poisson-like behavior; values above~$10^{2}-10^{3}$ signal dominance by rare extremes.
The sparsity and heavy tails that define event-driven telemetry data---low means from idle stretches, high variance from extreme bursts---inflate VMR far beyond mild-regime benchmarks.
Figure~\ref{fig:fano_ccdf} plots the distribution of per-entity VMR for each dataset; each dataset comprises hundreds to thousands of independently monitored entities, making the distribution informative.
Table~\ref{tab:temporal_baselines} reports median Mean Absolute Percentage Error (MAPE) for eight baselines across five mild-regime benchmarks (top) and seven wild-regime telemetry datasets (bottom), with per-dataset median VMR in the rightmost column.
A clear empirical gap separates the two regimes in our datasets: all benchmark median VMR values fall below~36, while all wild-regime telemetry median VMR values exceed~60.
Models designed for mild-regime data achieve MAPE below~$1$ on benchmarks but produce up to $53\times$ larger errors on wild-regime telemetry---even under the most conservative comparison (bottom row of Table~\ref{tab:temporal_baselines}).
\S\ref{subsec:forecasting} examines how sparsity and burstiness interact across datasets and granularities, and evaluates two mild-regime latency datasets as a sanity check confirming that \sys remains competitive outside its target regime.

\smartparagraph{Problem 2: A training objective mismatch.}
The data mismatch above explains why existing models fail at \emph{prediction}.
One baseline in Table~\ref{tab:temporal_baselines} (S2P2~\cite{chang2025s2p2}) achieves median MAPE of exactly~$1.0$ on every wild-regime dataset---it learns to predict zero, the safest output when most time bins are empty.
The \emph{representations} of this and similar models fail for a different reason: their training objective rewards accurate forecasts but places no specific constraints on the representation's geometric properties---how the model organizes its internal summaries~\cite{guthula_netfound_2025, beltiukov2025demystifying}.
Other baselines forecast non-trivially---N-BEATS~\cite{n-beats} achieves MAPE as low as~$0.84$ on some telemetry datasets---yet their embeddings concentrate all variance into a single dimension, collapsing behaviorally distinct time series into indistinguishable points (\S\ref{subsec:repr_quality}).
An operator asking ``what kind of traffic is this?'' receives a mixture of unrelated behaviors, and a search for ``when did I last see this pattern?'' returns largely arbitrary neighbors.

\S\ref{sec:design} presents a pipeline that addresses both problems.
\section{Design and Implementation}
\label{sec:design}

The core insight behind \sys is that disaggregating each time series into its constituent event streams---inter-burst gaps capturing burst timing and burst intensities capturing their magnitude---converts a single intractable wild-regime problem into two simpler streams, each resembling the dense, periodic data where existing models already succeed (\S\ref{sec:motivation}).
But the two streams are not independent---burst timing and magnitude are correlated---so a shared model must capture their dependencies.
The resulting representations must then serve the three operational tasks identified in \S\ref{sec:introduction}: forecasting, diagnosis, and semantic search.

\begin{figure*}[t]
    \centering
    \includegraphics[width=\linewidth]{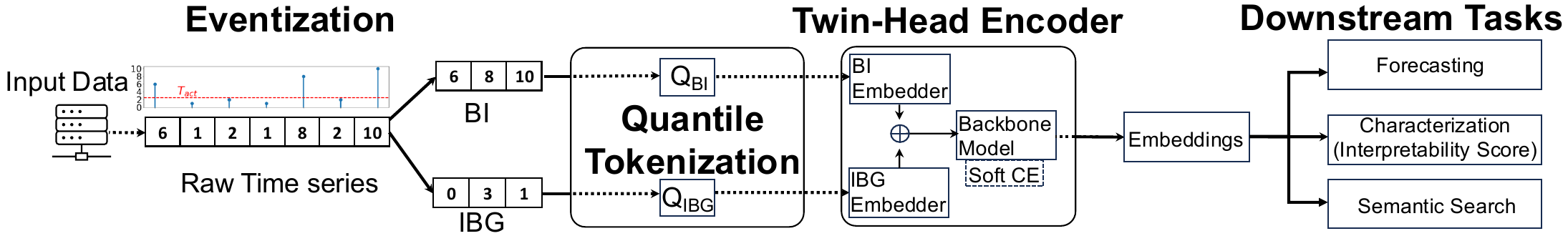}
    \caption{{\sys pipeline.} Raw telemetry is eventized into inter-burst gaps (IBG) and burst intensities (BI), discretized with quantile tokenizers, and encoded by a twin-head encoder. The resulting embedding supports forecasting, characterization, and search. }
    \label{fig:netBurstOverview}
    \vspace{-1em}
\end{figure*}

\sys realizes this as a two-phase pipeline (Figure~\ref{fig:netBurstOverview}).
The first phase contributes a novel event-centric pipeline where each stage addresses a specific consequence of the wild-regime structure (\S\ref{sec:representation}): eventization removes context waste, quantile tokenization preserves heavy tails, and a twin-head encoder with a soft cross-entropy objective produces well-spread embeddings that partition into operator-readable behavioral clusters.
The second phase applies these embeddings to forecasting, behavioral characterization, and semantic search (\S\ref{sec:downstream_design}).

\subsection{Learning Event-Centric Representations}
\label{sec:representation}

\smartparagraph{Eventization.}
Operationally irrelevant noise---including silence---dominates telemetry time series (\S\ref{sec:motivation}); eventization removes this noise.
Given a time series $s_{1:T}$ and a window length $w_{len}$, an activity threshold $T_{\mathrm{act}}$ declares a \emph{burst} whenever the aggregate value in a window exceeds this threshold.
Each burst at time $\tau_k$ yields two values: the \emph{inter-burst gap} $\mathrm{IBG}_k=\tau_k-\tau_{k-1}$ (with $\mathrm{IBG}_1=\tau_1$) capturing \textbf{when} the burst occurs, and the \emph{burst intensity} $\mathrm{BI}_k$ (the aggregate value over the burst window) capturing \textbf{how large} it is.
This transformation collapses long idle stretches into a single large IBG and replaces each burst with a compact BI value, producing two streams far shorter than the raw time series.
The two streams have distinct statistical structure---IBG captures incremental timing; BI isolates heavy-tailed burst magnitudes---and each individually resembles the milder data where existing models succeed, converting one wild-regime problem into two tractable ones.

$T_{\mathrm{act}}$ is an operator-controlled parameter---not a model hyperparameter---that defines the boundary between noise and operationally meaningful activity.
When the operator has no domain-specific preference, we provide a data-driven default: sweep percentiles of non-zero values and choose the lowest percentile where the VMR distribution first deviates from the unthresholded baseline (50th--80th percentiles across our datasets).
This selection of $T_{\mathrm{act}}$ strives to separate signal from noise without discarding operationally meaningful events.

\smartparagraph{Quantile tokenization.}
Existing models discretize values with uniform bins that undersample the rare extremes operators care about (\S\ref{sec:motivation}); quantile tokenization addresses this deficiency.
\sys replaces uniform binning~\cite{ansari2024chronos} with \emph{quantile-based codebooks}: \sys discretizes each stream so that every token corresponds to an equal fraction of probability mass---giving rare extreme values the same resolution as common ones.
\sys constructs global quantile codebooks $Q^{\text{IBG}}, Q^{\text{BI}}$ per dataset with fixed quantile boundaries; unlike learned codebooks~\cite{VQ-VAE}, the cut points remain fixed and reproducible across datasets, while the encoder learns the embedding vectors.

Both thresholding and quantization operate at \emph{dataset} granularity (e.g., per row in Table~\ref{tab:dataset_master}): one $T_{\mathrm{act}}$ and one set of quantile boundaries apply to all entities or time series in a dataset.
In contrast, Chronos-2~\cite{ansari2025chronos2} calibrates per individual time series, yielding tighter per-entity resolution but requiring a full pass over each entity's history---prohibitive at deployment scale at the scale of our datasets (Table~\ref{tab:dataset_master}).
We prefer quantile tokenization over log transforms because the latter introduce reconstruction bias when forecasts are mapped back to the original scale~\cite{biasCorrection,biasCorrectionReconciliation}.
At training time, the tokenizer maps each IBG or BI value to a token index $z^{\text{IBG}}_k, z^{\text{BI}}_k \in \{1,\dots,B\}$ according to its quantile bin ($B=4{,}096$ bins, following~\cite{ansari2024chronos}).
Forecasting then reduces to next-token prediction, and reconstruction replaces each token with its bin centroid. 
\S\ref{subsec:ablation} isolates the contribution of quantile tokenization by comparing against uniform binning on the same architecture.

\smartparagraph{Twin-head encoder.}
A shared encoder backbone with a separate prediction head per stream (\emph{twin-head encoder}) resolves the tension identified above: the backbone captures cross-stream dependencies through shared parameters, while the two heads preserve disentanglement at the output (\S\ref{subsec:ablation}).

\noindent\underline{\textit{Architecture.}}
The encoder takes the two quantized event streams (token sequences) from the previous stage as parallel inputs, each represented as a learned vector.
It fuses them into a single shared representation so that patterns in one stream can inform predictions in the other.
A shared T5 decoder~\cite{raffel_exploring_2020} processes this fused representation, while two separate prediction heads produce stream-specific forecasts: $p_\theta(z^{\text{IBG}}_k \mid h_k)$ for the next IBG value and $p_\psi(z^{\text{BI}}_k \mid h_k)$ for the next BI value, where $h_k$ is the decoder's hidden state at position $k$.

\noindent\underline{\textit{Soft cross-entropy.}}
Standard next-token cross-entropy penalizes every misprediction equally regardless of how close the predicted value is to the true one, concentrating representations into narrow regions where behaviorally different time series become indistinguishable~\cite{beltiukov2025demystifying}.
Distance-aware soft labels---where nearby classes receive more probability mass than distant ones---have proven effective for ordinal regression~\cite{diaz2019sord, imani2018distributional} and regression-via-classification~\cite{farebrother2024stopregressing}; Chronos~\cite{ansari2024chronos} identifies this as a promising but unexplored direction for time-series tokenization.
\sys applies this idea to quantile-tokenized telemetry, replacing standard targets with a \emph{soft cross-entropy} objective that spreads probability mass across neighboring quantile bins.
Unlike uniform label smoothing~\cite{szegedy2016labelsmoothing}, which distributes mass equally across all bins, this weighting respects the value ordering of quantile bins: bins representing similar values receive proportionally more mass, and a learnable per-stream sharpness $\alpha$ adapts to the distinct scales of IBG and BI.

Concretely, the loss for each position $k$ is
$\mathcal{L} = \sum_k -\sum_b q_b \log p_\theta(b \mid h_k)$, where $q_b \propto \exp\!\bigl(-\alpha \cdot |v_b - v_{z_k^*}| / \max(|v_{z_k^*}|, \epsilon)\bigr)$ weights each candidate bin $b$ by its value distance from the correct bin $z_k^*$, $h_k$ is the decoder's hidden state at position $k$, and the floor $\epsilon$ reflects each stream's value scale ($\epsilon_{\text{BI}} = 10^{-3}$, $\epsilon_{\text{IBG}} = 1.0$).
Training minimizes $\mathcal{L}_{\text{IBG}} + \mathcal{L}_{\text{BI}}$; at inference, each head generates tokens autoregressively and maps them back to values through the quantile codebooks.
\S\ref{subsec:repr_quality} validates that this objective produces the targeted well-spread geometry.

\subsection{Downstream Capabilities}
\label{sec:downstream_design}
\label{sec:rep_quality_design}

The shared representations generated by \sys supports three different tasks.
Each consumes the well-spread embedding geometry from \S\ref{sec:representation} and addresses one of the three operational needs from \S\ref{sec:introduction}: prediction, diagnosis, and search.

\smartparagraph{Forecasting via reconstruction.}
\label{sec:forecasting}
Forecasting predicts future events in the IBG and BI streams for any monitored entity---whether anticipating link saturation, capacity shortfalls, or traffic shifts.
Predicted IBGs accumulate into event times $\hat{\tau}_k = \hat{\tau}_{k-1} + \hat{\mathrm{IBG}}_k$, paired with predicted $\hat{\mathrm{BI}}_k$, recovering a time series that preserves burst statistics.
We evaluate forecasting with two complementary metrics: one for burst magnitude accuracy (MAPE) and one for temporal fidelity (Wasserstein distance, which measures the cost of aligning two distributions in time).
Because eventization separates timing from magnitude, a model can predict magnitudes correctly while misplacing bursts in time, or match the overall distribution while missing individual peaks---neither metric alone captures both failure modes.
\S\ref{subsec:forecasting} defines both metrics and reports results across all datasets.

\smartparagraph{Behavioral characterization via feature selection.}
\label{sec:characterization}
Characterization answers the operator's question about any cluster of time series: \emph{what kind of traffic is this?}
The answer must be in the statistical vocabulary operators already use---burstiness, periodicity, tail weight, complexity, etc.
TSFresh~\cite{tsfresh} defines this vocabulary with ${\sim}770$ hand-engineered statistics per time series.\footnote{Note that TSFresh features are not suited for clustering or search: their high-dimensional similarity conflates behaviorally distinct time series (\S\ref{subsec:repr_quality}), and their geometry collapses under search (\S\ref{subsec:search}).}
The question is which of these features best describe a given cluster.

The interpretable clustering literature identifies two criteria that a good descriptive feature must satisfy~\cite{hu2024interpretable}: \emph{alignment} (the embedding captures the feature within the cluster) and \emph{discrimination} (the feature distinguishes this cluster from others).
Alignment without discrimination produces generic descriptions that apply to every cluster; discrimination without alignment produces descriptions the embedding never learned to encode.
Existing methods instantiate these criteria with discrete structures---decision trees~\cite{moshkovitz2020explainable}, rule mining, learned gates~\cite{svirsky2024interpretable}---that scale poorly to high cluster counts.
Scaling interpretable clustering to hundreds of clusters remains an open problem~\cite{hu2024interpretable}.
\sys combines two continuous measures---CKA for alignment, Cohen's $\tilde{d}$ for discrimination---that sidestep discrete structures.
Local CKA (centered kernel alignment)~\cite{cka_similarity} quantifies alignment: within each cluster, it measures whether entities that are close in the embedding are also correlated to a candidate TSFresh feature, producing a score in $[0,1]$ regardless of feature scale.
Normalized Cohen's $\tilde{d}$~\cite{lakens2013effect, dalmaijer2022power} quantifies discrimination: for each feature $f$ in cluster $c$, standard Cohen's $d$ measures how many pooled standard deviations separate the cluster's mean from the global mean on that feature.
We normalize by the largest $|d|$ across all (feature, cluster)-pairs, mapping every score to $[0,1]$ so it is directly comparable to CKA.

\sys then ranks features by the product of both criteria: $\mathrm{CKA}_{c,f} \cdot \tilde{d}_{c,f} \in [0,1]$ for each cluster $c$ and feature $f$.
A cluster's \emph{interpretability score} sums the top-$j$ products:
\vspace{-0.25em}
\[
I_c \;=\; \sum_{f \in \text{top-}j}\; \mathrm{CKA}_{c,f}\cdot \tilde{d}_{c,f}\;,
\]
a scalar in $[0,j]$ summarizing how confidently the embedding admits an operator-readable description.
Both factors lie in $[0,1]$ regardless of feature scale or embedding dimensionality, so $I_c$ is directly comparable across embedding spaces.
This product-based scoring is new: it instantiates the established alignment-discrimination criteria~\cite{hu2024interpretable} with continuous measures from the statistics literature~\cite{lakens2013effect, dalmaijer2022power, cka_similarity}, capturing both alignment and discrimination while avoiding the discrete structures that limit existing methods to small cluster counts.
\S\ref{subsec:characterization} shows that \sys's $I_c$ distribution dominates baselines.

\smartparagraph{Semantic search.}
\label{sec:search}
Search retrieves historical time series whose behavior matches a given query---supporting incident response, root-cause analysis, and pattern-based triage.
Efficiency derives from two sources.
Eventization compresses each time series from thousands of raw timesteps to tens of event tokens, reducing per-entity embedding extraction cost proportionally.
Well-spread embeddings yield balanced partitions, bounding per-query retrieval to approximately $N/K$ vectors (where $N$ is the corpus size) rather than degrading toward brute-force scans.
Retrieval is also operationally meaningful: because embedding distances reflect behavioral similarity, nearest neighbors share operational characteristics by construction.
\S\ref{subsec:search} evaluates end-to-end search performance.

\subsection{Implementation}
\label{sec:implementation}

\sys is implemented in Python using PyTorch.
The twin-head encoder uses a T5-based transformer backbone (hidden size 512, 56.13M parameters) with two embedding layers fused via concatenation, layer normalization, and a feed-forward layer before a shared decoder.
Quantile codebooks use $B=4{,}096$ bins per stream.
Training uses Adam with early stopping on four NVIDIA A100 GPUs (12 wall-clock hours each, 48 GPU-hours total per dataset, 35.7 GB peak memory).
The encoder produces one 512-dimensional embedding per time series from the decoder hidden state; embedding extraction runs at 3{,}042 series/sec.
The characterization pipeline computes TSFresh features ($\sim$770 per time series) and the interpretability score defined in \S\ref{sec:characterization} using scikit-learn~\cite{pedregosa_scikit-learn_2011}.
The search index uses FAISS~\cite{faiss} with IVF-FLAT indexing and cosine similarity.
Clustering uses $K$-means~\cite{lloyd1982least} with $K$-means++ initialization~\cite{arthur2007kmeans}; its simplicity isolates the representation's contribution from the clustering algorithm.

\section{Evaluation}
\label{sec:evaluation}

Our evaluation proceeds in two phases. First, \S\ref{subsec:forecasting} compares \sys head-to-head against baselines on forecasting across all datasets, granularities, and burstiness regimes. Second, \S\ref{subsec:repr_quality}--\ref{subsec:search} examine representation quality, behavioral characterization, and semantic search on a single per-IP corpus---the geometric properties that forecasting alone cannot validate (\S\ref{sec:design}). \S\ref{subsec:ablation} ablates individual design choices.

\subsection{Experimental Setup}
\label{subsec:setup}

\begin{table}[t]
  \caption{Dataset summary. Columns follow the \S\ref{subsec:setup} pipeline: aggregation (Gran.) produces raw time series of length $TS_{Len}$, eventization at threshold $T_{\mathrm{act}}$ (per-dataset, not per-entity) yields $\#Events$, and VMR characterizes the resulting variability. $TS_{Len}$ and $\#Events$ report median and max across entities. $|D|$ reports total post-filter entities and the 30\% test partition used for forecasting evaluation (in thousands). $^{*}$Campus-Pkts-2 is not used for forecasting; $|D|$ reports all 54k post-filter entities, used for representation quality (\S\ref{subsec:repr_quality}--\S\ref{subsec:characterization}); search (\S\ref{subsec:search}) applies a separate 70:30 entity split.}
  \label{tab:dataset_master}
  \centering
  \resizebox{\columnwidth}{!}{%
  \begin{tabular}{l|rrrrrr}
    & \textbf{Gran.} & $\mathbf{TS_{Len}}$ & $\mathbf{T_{\mathrm{act}}}$ & $\mathbf{\#Events}$ & $\mathbf{|D|}$ ($\mathbf{\times 10^3}$) & \textbf{VMR} \\
    & & \textit{med\,/\,max} & & \textit{med\,/\,max} & \textit{tot\,/\,test} & \\
    \hline
Campus-Pkts-1 IP      & 1s    & 62 / 902      & 100\,B    & 7 / 902       & 512.4 / 153.6   & 1.2k  \\
Campus-Pkts-1 /24     & 1s    & 52 / 902      & 100\,B    & 8 / 902       & 269.3 / 80.7   & 1.3k \\
Campus-Pkts-1 Service & 100ms & 59 / 7{,}915  & 0\,B      & 8 / 6{,}969   & 284.7 / 85.2   & 102  \\
Campus-Pkts-2 IP      & 1s    & 60 / 902      & 100\,B    & 49 / 902      & 54.1$^{*}$     & 16.2k  \\
MAWI IP               & 1s    & 84 / 901      & 100\,B    & 5 / 901       & 10.8 / 3.3     & 307  \\
MAWI /24              & 1s    & 113 / 901     & 100\,B    & 11 / 901      & 11.0 / 3.3     & 1.1k  \\
MAWI Service          & 100ms & 71 / 8{,}979  & 0\,B      & 4 / 8{,}960   & 12.8 / 3.8     & 68  \\
RnE-NetFlow IP        & 60s   & 19 / 8{,}994  & 2.14\,MB  & 14 / 8{,}988  & 89.2 / 26.8    & 59.0M \\
RnE-Latency           & 1min  & 500 / 500     & 3\,ms     & 500 / 500     & 257.1 / 77.3   & 0.004  \\
Campus-Latency        & 3min  & 480 / 480     & 4\,ms     & 65 / 197      & 51.6 / 15.4    & 4.6  \\
  \end{tabular}%
  }
  \vspace{-1em}
\end{table}

\smartparagraph{Datasets.}
Table~\ref{tab:dataset_master} summarizes all ten dataset configurations. \noindent\emph{\underline{Volumetric datasets:}} Campus-Pkts-1 and Campus-Pkts-2 originate from the same campus gateway traces (15-minute capture windows); Campus-Pkts-1 serves forecasting at service, IP, and subnet granularities, Campus-Pkts-2 preserves per-IP granularity for clustering and search, filtered to a specific link-capacity regime (1\,MB minimum transfer, 6\,Mbps maximum throughput) that removes both inactive mice and outlier elephants. This filtering produces a denser dataset than Campus-Pkts-1 (median 49 events out of 60 steps), placing it closer to the mild regime where \sys's eventization advantage is smallest---a conservative choice for evaluating representation quality. MAWI~\cite{cho_traffic_2000} provides ISP backbone traces; RnE-NetFlow captures R\&E backbone traffic via production NetFlow~\cite{claise_cisco_2004} collectors. 

\noindent\emph{\underline{Latency datasets:}} RnE-Latency reports 95th-percentile latency from production perfSONAR probes~\cite{tierney_perfsonar_2009} between R\&E backbone node pairs (1-minute granularity); Campus-Latency collects ICMP round-trip times across campus measurement nodes (3-minute granularity). The two latency datasets inhabit the mild regime (VMR~$\leq 4.6$), serving as a control: they test whether \sys's event-centric design degrades when the data lacks the sparse, bursty structure it targets.

\smartparagraph{Aggregation.}
For unsampled packet traces (Campus-Pkts-1, Campus-Pkts-2, MAWI), we choose the \emph{temporal aggregation} granularity: 100\,ms for service-level traces (fine enough for QoS monitoring and anomaly detection; 1\,s is too coarse to capture individual burst events) and 1\,s for IP- and subnet-level traces (100\,ms at IP granularity produces noisy, operationally uninformative time series). For RnE-NetFlow and the latency datasets, the sampling or reporting frequency dictates the granularity: 60\,s for RnE-NetFlow (90\% of inter-record gaps fall within 60\,s; finer granularity injects artificial zeros). \emph{Spatial aggregation} groups traffic at three granularities---service-level (source IP, destination IP, min of the two ports---grouping all flows between the same endpoints to the same well-known service), IP-level (per-endpoint), and subnet-level (/24, per-network-segment)---each revealing different operational patterns. \S\ref{sec:discussion} discusses the scope of this selection relative to other possible aggregations.

\smartparagraph{Eventization.}
\sys applies $T_{\mathrm{act}}$ from \S\ref{sec:representation} to each aggregated time series; Table~\ref{tab:dataset_master} reports per-dataset values (\S\ref{subsec:design_validation} confirms robustness to this choice).
We exclude entities with fewer than 3 events after eventization and truncate each time series at its last observed event---otherwise idle-tail entities reward models that predict silence. Evaluation partitions data along two orthogonal dimensions. \emph{Entity split:} 70\% of entities are reserved for training or fine-tuning; $|D|$ in Table~\ref{tab:dataset_master} reports the remaining 30\% used for forecasting evaluation. \emph{Temporal split:} within each test entity, the forecasting protocol (\S\ref{subsec:forecasting}) observes the first 70\% of time steps and predicts the remaining 30\%. The resulting corpora range from ${\sim}$3k--154k test entities per configuration.
Note that (1)~eventization and quantile codebook fitting ($4{,}096$ bins) use training data only to prevent leakage; and (2)~eventization applies only to \sys's input---all baselines receive raw aggregated time series, though both predicted and ground-truth series are thresholded before metric computation (\S\ref{subsec:forecasting}).
Representation quality and behavioral characterization (\S\ref{subsec:repr_quality}--\S\ref{subsec:characterization}) use all entities passing the 3-event minimum (e.g., 54K for Campus-Pkts-2). Semantic search (\S\ref{subsec:search}) applies a separate 70:30 entity-level split: 70\% for index construction, 30\% for retrieval queries.

\smartparagraph{Baselines and comparison targets.}
We compare against eight forecasting baselines spanning three categories: five time-series foundation models (Chronos~\cite{ansari2024chronos}, Chronos-2~\cite{ansari2025chronos2}, Lag-Llama~\cite{lagllama}, Toto~\cite{toto}, S2P2~\cite{chang2025s2p2}), two deep forecasters (DeepAR~\cite{deepar}, N-BEATS~\cite{n-beats}), and one transformer-based point-process model (THP~\cite{transformer_hawkes}). TSFresh~\cite{tsfresh} serves as an additional baseline for representation quality, characterization, and search (\S\ref{subsec:repr_quality}--\ref{subsec:search}), and provides the statistical feature vocabulary for cluster attribution (\S\ref{sec:characterization}).
All foundation models except Toto are fine-tuned from public checkpoints; Toto is evaluated zero-shot (fine-tuning code was not publicly available at the time of our experiments). Training details and learning-rate choices are in Appendix Table~\ref{tab:baselines}.
Unless specified otherwise, we select Chronos-2 as the primary representative baseline because it is the strongest forecasting competitor on wild-regime data (Table~\ref{tab:master_results})---making it the natural target for analyzing whether forecasting accuracy implies representation quality.

\vspace{-1em}

\subsection{Forecasting}
\label{subsec:forecasting}

\begin{table}[t]
  \caption{Forecasting results on matched entities. Median MAPE (primary: per-burst magnitude accuracy) and median WD (secondary: distributional fidelity at full forecast horizon). Parentheses: factor $\times$ improvement of \sys over strongest baseline ($>$1 = \sys wins). Strongest baseline selected by median MAPE.}
  \label{tab:master_results}
  \centering
  \resizebox{\columnwidth}{!}{%
  \begin{tabular}{l|llc}
    \textbf{Configuration} & \textbf{Median MAPE} & \textbf{Median WD} & \textbf{Best Baseline} \\
    \midrule
    Campus-Pkts-1 IP      & 0.565 (1.73$\times$)  & 0.408 (2.19$\times$)  & Chronos-2 \\
    Campus-Pkts-1 /24     & 0.600 (1.61$\times$)  & 0.418 (3.72$\times$)  & N-BEATS \\
    Campus-Pkts-1 Service & 0.008 (116$\times$)   & 0.110 (7.48$\times$)               & Toto \\
    MAWI IP               & 0.487 (1.98$\times$)  & 0.1303 (1.60$\times$)  & N-BEATS \\
    MAWI /24              & 0.521 (1.58$\times$)  & 0.525 (7.58$\times$)  & N-BEATS \\
    MAWI Service          & 0.009 (90.1$\times$)  & 0.103 (1.69$\times$)               & N-BEATS \\
    RnE-NetFlow IP        & 0.586 (1.27$\times$)  & 0.500 (1.04$\times$)               & Chronos-2 \\
    RnE-Latency           & 0.011 (1.19$\times$)  & 0.1862 (0.53$\times$)               & Toto \\
    Campus-Latency        & 0.195 (1.92$\times$)  & 5.138 (1.21$\times$)               & Chronos-2 \\
  \end{tabular}%
  }
  \vspace{-1em}
\end{table}

Forecasting is the most direct way to test whether the encoder captured burst structure. We evaluate across all nine forecasting configurations in Table~\ref{tab:dataset_master} (Campus-Pkts-2 is reserved for clustering and search) using two complementary metrics; Table~\ref{tab:master_results} reports results.

\smartparagraph{Success metrics.}
\emph{Mean Absolute Percentage Error (MAPE)} compares predicted and actual burst magnitudes on events only, excluding idle periods where a trivial zero predictor achieves deceptively low error. A baseline that predicts zero everywhere scores MAPE\,$=$\,1; \sys's MAPE of 0.008--0.60 across configurations confirms that it predicts bursts and approximates their magnitudes. MAPE has a known asymmetry against bold predictors: overshooting a small burst inflates MAPE above 1, while predicting zero is bounded at 1 regardless of the miss. The improvement ratios in Table~\ref{tab:master_results} are therefore a \emph{lower bound} on \sys's operational advantage.
\emph{Wasserstein distance (WD)} complements MAPE by capturing timing drift and false-positive errors. We normalize predicted and ground-truth time series to unit sum, treating each as a distribution over time steps, and compute the 1-Wasserstein distance~\cite{wasser}---the minimum cost of transporting mass from one temporal profile to the other. A burst predicted too early or too late incurs cost proportional to its displacement; a missing or spurious burst incurs cost for the mass that must be supplied or removed. Both MAPE and WD evaluate on the thresholded time series (events only); baselines receive the full raw time series as input (\S\ref{subsec:setup}) so no information is lost in their prediction, but both predicted and ground-truth series are thresholded before metric computation to focus evaluation on the operationally relevant signal.

\begin{figure}[t]
  \centering
  \includegraphics[width=\linewidth]{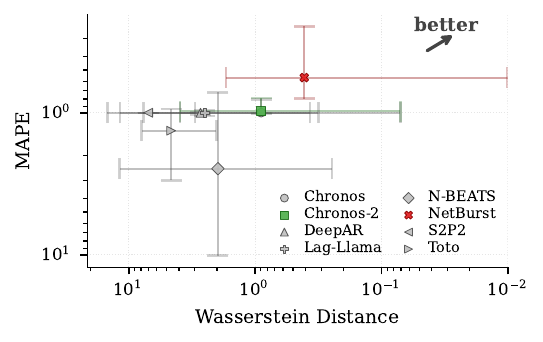}
  \caption[MAPE vs.\ WD frontier on Campus-Pkts-1 IP]{MAPE vs.\ WD frontier on Campus-Pkts-1 IP (1\,s). Each point is one model's median performance; axes are log-scaled with the origin at top-right (better).\footnotemark}
  \label{fig:frontier_mape_wd}
  \vspace{-1em}
\end{figure}
\footnotetext{THP is omitted because its WD is too high to display on the same scale.}

\smartparagraph{\sys leads across all configurations.}
Table~\ref{tab:master_results} summarizes median MAPE and WD across all nine forecasting configurations against each configuration's strongest competitor (selected by median MAPE). The table reports medians; full per-entity MAPE and WD distributions for all models appear in Appendix~\ref{sec:appendix_forecasting_distributions}. Figure~\ref{fig:frontier_mape_wd} complements the table by showing \emph{all} models simultaneously on Campus-Pkts-1 IP: while the table compresses each configuration to a single ratio, the frontier plot reveals the competitive landscape---\sys occupies the Pareto-optimal upper-right corner (lowest MAPE, lowest WD), while baselines scatter across the high-error lower-left region.

Wild-regime datasets---\sys's design target---see $1.3$--$116\times$ improved MAPE scores across both volumetric (Campus-Pkts-1, MAWI, RnE-NetFlow) and latency modalities. On mild-regime data (RnE-Latency), \sys stays competitive at $1.19\times$, confirming that event-centric design does not sacrifice performance outside the wild regime it targets. WD confirms distributional fidelity with $1.0$--$7.5\times$ improvement across wild-regime datasets. On RnE-Latency, \sys's WD is $1.9\times$ worse than Chronos-2: the dataset-wide activity threshold over-segments the relatively smooth latency signal, complicating burst prediction. Per-entity threshold calibration remains open (\S\ref{sec:limitations}).

\smartparagraph{Sparsity and burstiness together drive the largest gains.}
High VMR alone does not predict \sys's advantage---the largest gains appear where sparsity and burstiness coincide. The two service-level configurations---Campus-Pkts-1 and MAWI, with 116× and 90× gains respectively---illustrate this interaction: at 100~ms granularity, in the median Campus-Pkts-1 entity 86\% of time steps carry no operationally relevant signal, and 94\% in MAWI (Table~\ref{tab:dataset_master}); eventization (\S\ref{subsec:setup}) compresses these to 8 and 4 events respectively, while baselines process the full trace. At the IP and subnet levels (1~s, VMR up to $1.2$k), the improvement is $1.6$--$2.0\times$--- the baselines still produce 60--100\% more error than \sys, but lower sparsity lets them retain more signal per time step. Binning entities by VMR on Campus-Pkts-1 IP confirms this: \sys produces lower MAPE than Chronos-2 on ${\approx}$93\% of entities at VMR,$<$,$10^{2}$ and ${\approx}$86\% at VMR,$>$,$10^{3}$---a roughly flat win rate across burstiness, confirming that the gain comes from input
  decomposition, not from the statistical regime itself.

\smartparagraph{Takeaway.}
\sys dominates all baselines on forecasting accuracy (Table~\ref{tab:master_results}, Figure~\ref{fig:frontier_mape_wd}) and distributional fidelity (Table~\ref{tab:master_results}). However, forecasting accuracy alone does not guarantee representation quality---we show next that a strong forecaster like N-BEATS can still produce collapsed representations.

\subsection{Representation Quality}
\label{subsec:repr_quality}

\begin{table}[t]
  \caption{Representation quality on Campus-Pkts-2 /32 IP. Geometric quality measures embedding capacity; feature alignment (CKA) measures correlation with TSFresh statistics. $\downarrow$\,=\,lower is better; $\uparrow$\,=\,higher is better.}
  \label{tab:repr_geometry}
  \centering
  \resizebox{.9\columnwidth}{!}{%
    \begin{tabular}{l|ccc|c}
      & \multicolumn{3}{c|}{\textbf{Geometric quality}} & \textbf{Alignment} \\
      Model & Cos$\downarrow$ & MCC$\downarrow$ & Top E/V$\downarrow$ & CKA$\uparrow$ \\
      \hline
      \sys       & \textbf{0.500} & \textbf{0.106} & \textbf{0.206} & \textbf{0.23} \\
      Chronos-2~\cite{ansari2025chronos2}   & 0.699 & 0.226 & 0.273 & 0.13 \\
      Chronos~\cite{ansari2024chronos}    & 0.623 & 0.293 & 0.197 & 0.09 \\
      Toto~\cite{toto}       & 0.788 & 0.470 & 0.412 & 0.10 \\
      N-BEATS~\cite{n-beats}    & 0.004 & 0.003 & 0.999 & 0.01 \\
      TSFresh~\cite{tsfresh}    & 0.652 & 0.687 & 1.000 & 1.00 \\
    \end{tabular}
    }
\end{table}

Representation geometry determines whether clustering and search succeed; forecasting alone cannot test it.
We evaluate this claim on all 54K post-filter Campus-Pkts-2 IP entities (Table~\ref{tab:dataset_master}), where the conservative filtering of \S\ref{subsec:setup} produces a milder regime---denser data with less eventization advantage---making this a stricter case for testing representation quality.
Table~\ref{tab:repr_geometry} quantifies two complementary dimensions: \emph{geometric quality}---whether the embedding uses its full capacity rather than collapsing~\cite{beltiukov2025demystifying}---and \emph{feature alignment}---whether learned representations correlate with operator-recognizable statistics such as burstiness and periodicity.

\smartparagraph{Geometric quality.}
Three complementary diagnostics capture different failure modes of embedding geometry (Table~\ref{tab:repr_geometry}, left columns).
\emph{Mean pairwise cosine} (\emph{Cos}) measures angular spread: 0 means vectors are evenly distributed; values near~1 indicate ``cone collapse,'' where all embeddings point in the same direction~\cite{beltiukov2025demystifying}.
\emph{Max cosine contribution} (\emph{MCC}) measures directional concentration: the fraction of total pairwise similarity explained by the single most aligned dimension. High MCC means one axis dominates all distances, so the embedding ignores variation along every other behavioral dimension.
\emph{Top explained variance} (\emph{Top E/V}) measures spectral concentration: the fraction of total variance captured by the first principal component. High Top~E/V means the embedding is effectively one-dimensional.
Each baseline fails on at least one diagnostic: Chronos-2 and Toto exhibit angular cone collapse (Cos 0.70 and 0.79); N-BEATS collapses dimensionally (Top~E/V\,=\,0.999), concentrating all variance into a single principal component despite low Cos. \sys achieves the strongest scores on all three diagnostics.
Chronos-2 serves as the primary representation baseline throughout \S\ref{subsec:characterization}--\ref{subsec:search} because it is the strongest forecaster on wild-regime data (Table~\ref{tab:temporal_baselines}) with better geometry than Toto (Cos 0.70 vs.\ 0.79, Top~E/V 0.273 vs.\ 0.412).

\smartparagraph{Feature alignment.}
Geometric quality ensures distances are meaningful, but an operator also needs to \emph{describe} what a cluster represents.
We measure feature alignment with linear CKA~\cite{cka_similarity} (centered kernel alignment), a scale-invariant correlation between the 512-D embedding space and the 770-D TSFresh~\cite{tsfresh} feature space (Table~\ref{tab:repr_geometry}, right column). \sys's CKA of 0.23 is the highest among representation learners, confirming that its embeddings capture operator-recognizable statistics such as burstiness, periodicity, and tail weight.
TSFresh achieves CKA\,=\,1.00 by construction, but this alignment is illusory: its dimensional collapse (Top~E/V\,=\,1.000) reduces 770 features to one effective dimension. N-BEATS exhibits similar dimensional collapse (Top~E/V\,=\,0.999) with no interpretable structure (CKA\,=\,0.01). \sys is the only model that achieves both geometric spread and feature alignment.

\begin{figure}[t]
  \centering
  \begin{subfigure}[t]{0.48\linewidth}
      \centering
      \includegraphics[width=\linewidth]{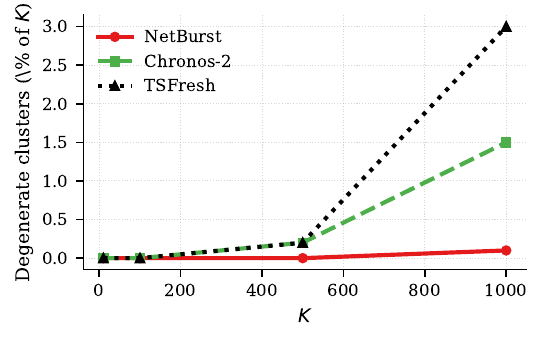}
      \caption{Degeneracy vs $K$}
      \label{fig:degeneracy_vs_k}
  \end{subfigure}\hfill
  \begin{subfigure}[t]{0.48\linewidth}
      \centering
      \includegraphics[width=\linewidth]{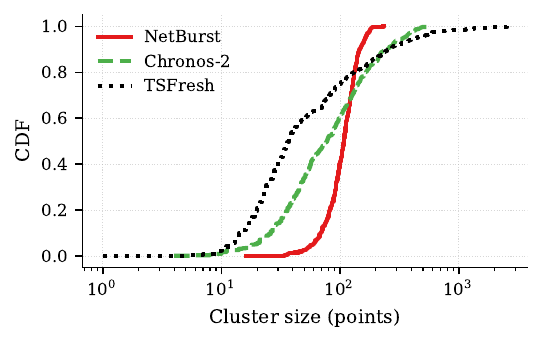}
      \caption{Cluster size CDF ($K\!=\!500$)}
      \label{fig:cluster_size_cdf}
  \end{subfigure}
  \caption{Cluster quality under $K$-means. (a)~Degeneracy vs.\ $K$. (b)~Cluster size CDF at $K\!=\!500$.}
  \label{fig:repr_quality}
\end{figure}

\smartparagraph{Cluster quality follows from the embedding.}
Using the $K$-means clustering described in \S\ref{sec:implementation}, we measure two properties of the resulting partitions: \emph{degeneracy} (fraction of clusters with fewer than 20 members~\cite{dalmaijer2022power}) and \emph{balance} (portion of clusters of different sizes). Figure~\ref{fig:repr_quality}(a) shows that \sys produces near-zero degenerate clusters as $K$ sweeps from 100 to 1,000; baselines degenerate beyond 500 clusters. Figure~\ref{fig:repr_quality}(b) shows the cluster size distribution at $K\!=\!500$: \sys is tightly concentrated, TSFresh has a heavy tail of oversized clusters. Normalized Shannon entropy $H_{\text{norm}} = H / \log K$ quantifies balance as a single scalar ($1.0$ = all clusters equal size; $0$ = all points in one cluster). \sys achieves $H_{\text{norm}} = 0.993$ vs.\ 0.943 (Chronos-2) vs.\ 0.853 (TSFresh). Balanced clusters bound per-query search cost to ${\sim}N/K$ (\S\ref{subsec:search}).

\smartparagraph{Takeaway.}
\sys dominates all baselines on geometric quality (Table~\ref{tab:repr_geometry}), feature alignment (Table~\ref{tab:repr_geometry}), and cluster quality (Figure~\ref{fig:repr_quality}).

\begin{figure}[h]
  \centering
  \includegraphics[width=\linewidth]{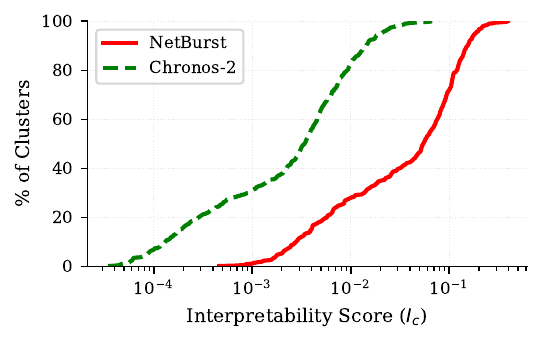}
  \caption{CDF of the per-cluster interpretability score $I_c$ on Campus-Pkts-2 /32 IP ($K\!=\!500$).}
  \label{fig:cluster_quality_cdf}
  \vspace{-2em}
\end{figure}

\begin{table*}[t]
  \caption{Natural-language descriptions of five \sys clusters sampled across the $I_c$ distribution ($K\!=\!500$, Campus-Pkts-2 /32 IP): ranks 1--2 (highest $I_c$), rank $\approx\!250$ (median), and ranks 499--500 (lowest $I_c$). Per-cluster top-5 features ranked by $\mathrm{CKA}\cdot\tilde{d}$ with cluster means. Feature notation is defined below.$^{\dagger}$}
  \begin{minipage}{\textwidth}
  \vspace{2pt}
  \footnotesize $^{\dagger}$Unless prefixed, each feature is \texttt{change\_quantiles}: variance of step-to-step changes restricted to values whose percentile rank falls in $[q_a,q_b]$. ``$\pm$'' = signed; ``$|\cdot|$'' = absolute. Other prefixes: \texttt{cq-mean} = mean of changes; \texttt{q} = plain quantile; \texttt{AR} = AR coefficient (lag~10); \texttt{CV} = coefficient of variation; \texttt{psd} = spectral density (band~8).
  \end{minipage}
  \label{tab:cluster_taxonomy}
  \centering
  \begin{small}
  \setlength{\tabcolsep}{4pt}
  \renewcommand{\arraystretch}{1.1}
  \begin{tabular}{@{}c>{\raggedright\arraybackslash}p{1.8cm}>{\raggedright\arraybackslash}p{5.0cm}>{\raggedright\arraybackslash}p{9.0cm}@{}}
  \textbf{Rank} & \textbf{Score ($I_c$)} & \textbf{Top-5 descriptive features} & \textbf{Natural language description} \\
  \midrule
  1   & 0.390
      & $\pm[.4,.6]{:}\,372$;\, $|\cdot|[.2,.6]{:}\,365$;\, $|\cdot|[.2,.4]{:}\,100$;\, $\pm[.0,.2]{:}\,137$;\, $|\cdot|[.0,.2]{:}\,99$
      & Frequent moderate bursts with rapid transitions; the largest step-to-step changes occur among mid-range values. These features capture volatility structure across the entity's operating range. \\
  2   & 0.381
      & $\pm[.0,.2]{:}\,172$;\, $\pm[.0,.4]{:}\,1156$;\, $|\cdot|[.0,.2]{:}\,101$;\, $|\cdot|[.0,.4]{:}\,556$;\, \texttt{q}$_{.4}{:}\,42.6$
      & Persistently active traffic that maintains a nonzero baseline, with high volatility concentrated among the lower-intensity values. These features capture activity persistence and the variability of transitions in the low-to-moderate range. \\
  250 & 0.054
      & $\pm[.0,.4]{:}\,3.1$;\, $|\cdot|[.0,.4]{:}\,2.7$;\, \texttt{q}$_{.4}{:}\,0.50$;\, \texttt{median}${:}\,10.2$;\, \texttt{kurt}${:}\,7.2$
      & Mostly idle traffic with a low typical level and occasional bursts that far exceed the baseline. These features capture the typical activity level, the prevalence of near-zero values, and the heaviness of the burst tail. \\
  499 & $6.8{\times}10^{-4}$
      & $|\cdot|[.2,.6]{:}\,3.4{\times}10^{5}$;\, $\pm[.2,.6]{:}\,6.8{\times}10^{5}$;\, \texttt{AR}${:}\,1308$;\, \texttt{CV}${:}\,3.4$;\, \texttt{psd}${:}\,5.6{\times}10^{8}$
      & Highly variable traffic with sustained burst episodes that persist across consecutive time steps and recur with detectable regularity. These features capture variability relative to the mean, temporal persistence, and spectral structure. \\
  500 & $4.7{\times}10^{-4}$
      & \texttt{cq-mean}$\,\pm[.0,.4]{:}\,0.23$;\, $\pm[.0,.4]{:}\,8.8{\times}10^{4}$;\, \texttt{AR}${:}\,1288$;\, \texttt{kurt}${:}\,20.4$;\, \texttt{max}${:}\,7.4{\times}10^{4}$
      & Rare extreme spikes on an otherwise quiet background, with episodes that build gradually rather than appearing instantaneously. These features capture tail heaviness, peak magnitude, and temporal dependence between consecutive values. \\
  \end{tabular}
  \end{small}
  \end{table*}
\begin{figure*}[!t]
    \centering
    \begin{subfigure}[t]{0.2\textwidth}
        \centering
        \includegraphics[width=\linewidth]{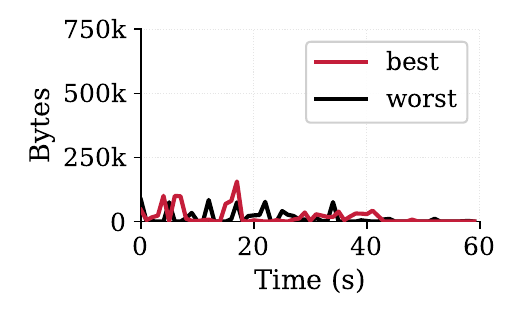}
        \caption{Rank 1}
        \label{fig:taxonomy_rank1}
    \end{subfigure}%
    \hfill
    \begin{subfigure}[t]{0.2\textwidth}
        \centering
        \includegraphics[width=\linewidth]{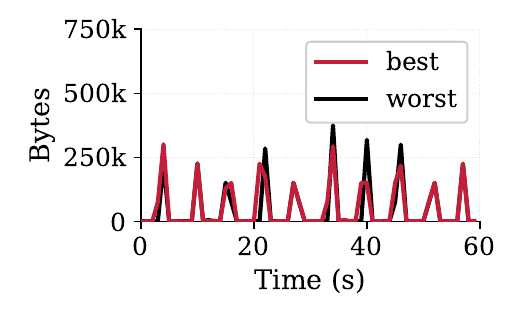}
        \caption{Rank 2}
        \label{fig:taxonomy_rank2}
    \end{subfigure}%
    \hfill
    \begin{subfigure}[t]{0.2\textwidth}
        \centering
        \includegraphics[width=\linewidth]{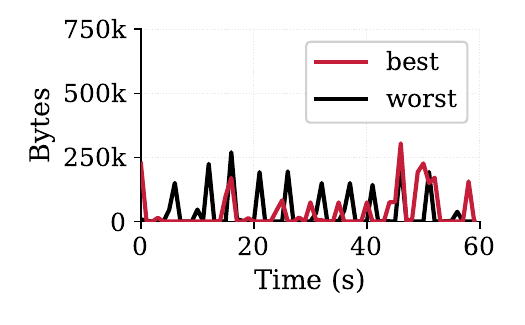}
        \caption{Rank 250}
        \label{fig:taxonomy_rank250}
    \end{subfigure}%
    \hfill
    \begin{subfigure}[t]{0.2\textwidth}
        \centering
        \includegraphics[width=\linewidth]{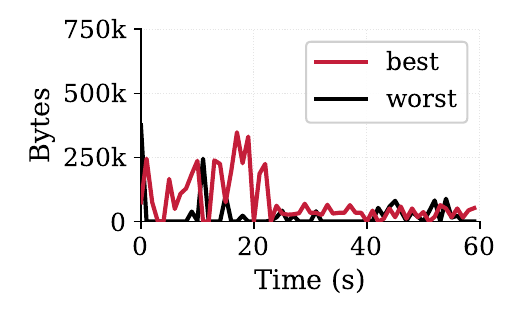}
        \caption{Rank 499}
        \label{fig:taxonomy_rank499}
    \end{subfigure}%
    \hfill
    \begin{subfigure}[t]{0.2\textwidth}
        \centering
        \includegraphics[width=\linewidth]{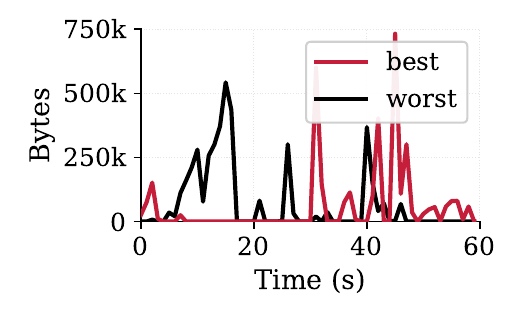}
        \caption{Rank 500}
        \label{fig:taxonomy_rank500a}
    \end{subfigure}%
    \caption{\sys cluster overlays: best (red) and worst (black) match to the cluster's feature profile among 10 random samples for each of the five clusters in Table~\ref{tab:cluster_taxonomy}.}
    \label{fig:cluster_examples}
    \vspace{-1em}
\end{figure*}

\subsection{Behavioral Characterization}
\label{subsec:characterization}

\S\ref{subsec:repr_quality} established embedding quality; we now test whether the resulting clusters are operationally meaningful. The interpretability score $I_c$ (\S\ref{sec:characterization}) quantifies how describable each cluster is. In particular, we want to know if the defining features of a cluster accurately describe what its members (i.e., time series) look like and whether higher $I_c$ values are a reflection of higher confidence in such descriptions of the obtained clusters.

\smartparagraph{\sys's $I_c$ distribution consistently exceeds Chronos-2.}
Figure~\ref{fig:cluster_quality_cdf} compares the CDF of $I_c$ across all $K\!=\!500$ clusters for \sys and Chronos-2 on Campus-Pkts-2 IP (same pipeline and features---only the embeddings differ). \sys scores $16\times$ higher at the median across the full distribution. The gap reflects not only higher CKA (alignment, consistent with the embedding geometry in \S\ref{subsec:forecasting}) but also higher Cohen's $\tilde{d}$ (discrimination)---\sys's top features distinguish each cluster from the corpus, not just correlate with it.

\smartparagraph{Cluster descriptions derived from features alone.}
Table~\ref{tab:cluster_taxonomy} samples five clusters across the $I_c$ spectrum (top two, median, bottom two) and lists each cluster's top-5 features ranked by $I_c$. We derive each natural-language description by prompting an LLM (Claude Opus~4.6~\cite{anthropic2025claude}) with only the feature names and their average values across cluster members---no raw time series are inspected. The same prompt is applied to every cluster; each description is regenerated 10 times and the most stable wording is selected to mitigate LLM output variability (Appendix~\ref{sec:llm_prompt}).
High-$I_c$ clusters draw on \emph{percentile corridors} (ranges $[q_a, q_b]$ restricting the measurement to a slice of the feature distribution), while lower-$I_c$ clusters draw on distribution-level properties such as coefficient of variation, autoregressive coefficients, and kurtosis.

\smartparagraph{Higher $I_c$ clusters have more representative feature profiles.}
For each cluster in Table~\ref{tab:cluster_taxonomy}, we draw 10 random members, rank them by how close their feature values are to the cluster average, and overlay the best match (red) and worst match (black) in Figure~\ref{fig:cluster_examples}.
At the high end of $I_c$ (ranks~1--2, Figure~\ref{fig:cluster_examples}a,b), the two members exhibit the behavioral pattern the features describe: rank~1's members show frequent moderate bursts with rapid transitions; rank~2's members show persistent, nearly continuous activity---matching the feature profile in Table~\ref{tab:cluster_taxonomy}.
At the low end of $I_c$ (ranks~499--500, Figure~\ref{fig:cluster_examples}d,e), the best match still matches the feature profile, but the worst match diverges visibly---rank~499's members share the same behavioral pattern---sustained active periods---but differ in intensity; the description covers both less precisely, which is what a low $I_c$ score predicts.

We repeat this analysis on Chronos-2 (Appendix Table~\ref{tab:selected_cluster_taxonomy_chronos2}, Figure~\ref{fig:cluster_examples_chronos2}).
In the case of Chronos-2's \emph{best} cluster ($I_c = 0.058$, rank~1, Figure~\ref{fig:cluster_examples_chronos2}a), the best match shows regular large spikes every few seconds---a clearly periodic burst pattern. The worst match is a nearly flat, low-amplitude signal with no discernible burst structure. The embedding produced by Chronos-2 placed these two fundamentally different behavioral types in the same cluster; no single description can cover both.
Chronos-2's rank~2 (Figure~\ref{fig:cluster_examples_chronos2}b) is the one case where both members do look alike---both show frequent moderate bursts---and the description matches what both traces show.
At rank~499 (Figure~\ref{fig:cluster_examples_chronos2}d), the pattern repeats: the best match shows periodic high-amplitude spikes while the worst match hovers near zero---the same categorical mismatch as rank~1.
In summary, Chronos-2's clusters mix behavioral types that \sys's clusters separate, and this failure appears at Chronos-2's top rank, not just at the bottom of the distribution.

\smartparagraph{Takeaway.}
$I_c$ quantifies a testable property: whether a cluster's defining features accurately describe its members. \sys maintains this agreement across its top ${\sim}250$ clusters (Figure~\ref{fig:cluster_quality_cdf}); Chronos-2's best cluster already mixes behavioral types that no single description can cover (Appendix Figure~\ref{fig:cluster_examples_chronos2}). A full calibration of $I_c$ against operator judgment is future work  (\S\ref{sec:discussion}).

\subsection{Semantic Search}
\label{subsec:search}

\S\ref{subsec:forecasting} validated forecasting accuracy, \S\ref{subsec:repr_quality} established balanced cluster geometry, and \S\ref{subsec:characterization} confirmed that clusters admit interpretable descriptions. None test whether these properties translate into operationally efficient retrieval. We index the 70\% training partition of Campus-Pkts-2 (${\sim}$38K embeddings) with FAISS~\cite{faiss} using IVF-FLAT ($K\!=\!500$ partitions, cosine similarity); the $\mathrm{nprobe}$ parameter controls how many partitions each query searches. Per-query latency has two components: \emph{embedding extraction}---computing a 512-d vector from a raw time series, governed by the input compression of \S\ref{sec:design}---and \emph{retrieval}---searching within the assigned IVF partitions, governed by the cluster balance of \S\ref{subsec:repr_quality}.

\begin{figure}[t]
    \centering
    \includegraphics[width=\linewidth]{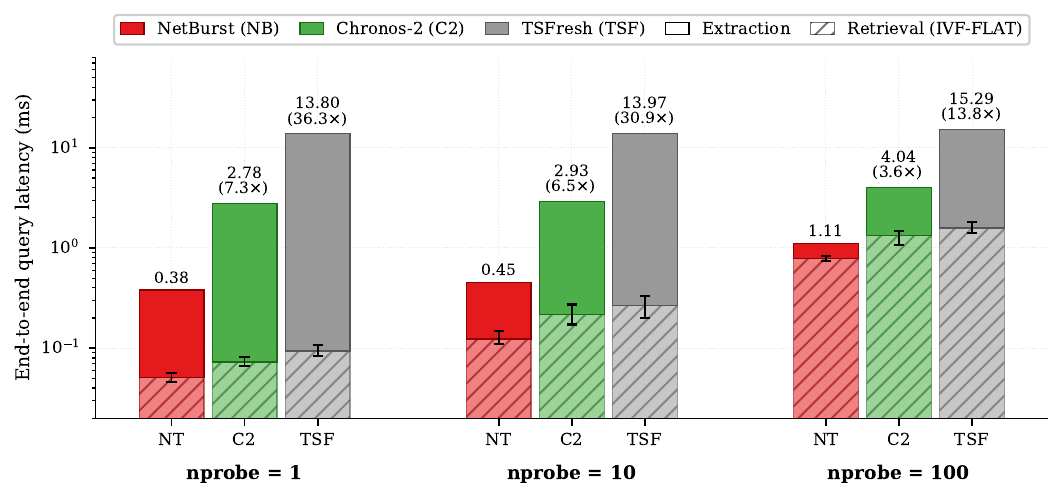}
    \caption{Per-query latency (log scale) at $\mathrm{nprobe} \in \{1, 10, 100\}$ on 50K Campus-Pkts-2 embeddings ($K\!=\!500$). Each bar stacks embedding extraction (solid, top) and IVF-FLAT retrieval (hatched, bottom); whiskers show retrieval $p_{25}$--$p_{75}$; totals and slowdown ratios vs.\ \sys annotated above each bar.}
    \label{fig:search_scalability}
    \vspace{-1em}
\end{figure}

\smartparagraph{\sys is faster at both extraction and retrieval.}
Figure~\ref{fig:search_scalability} shows both components at $\mathrm{nprobe} \in \{1, 10, 100\}$. Event-centric input compression (\S\ref{sec:design}) reduces extraction to $0.33$\,ms per vector---$8.2\times$ faster than Chronos-2 ($2.71$\,ms) and $50\times$ faster than TSFresh ($16.4$\,ms)---because \sys processes ${\sim}50$ event tokens per series where Chronos-2 processes 512 raw timesteps. Balanced partitions ($H_{\text{norm}}=0.993$) bound retrieval to ${\sim}N/K$ vectors per query, yielding tight $p_{25}$--$p_{75}$ whiskers and a $1.5$--$2\times$ retrieval advantage at every $\mathrm{nprobe}$. End-to-end, \sys is $7.5\times$ faster than Chronos-2 and $45\times$ faster than TSFresh at $\mathrm{nprobe}=1$. The gap narrows at higher $\mathrm{nprobe}$ ($3.8$--$17\times$ at $\mathrm{nprobe}=100$) because extraction cost is independent of $\mathrm{nprobe}$ while retrieval cost grows linearly with the number of partitions searched. Chronos-2 and TSFresh bars appear flat across $\mathrm{nprobe}$ on the log scale: extraction dominates their total cost, so partition-level refinements remain invisible end-to-end.

\smartparagraph{Takeaway.}
Compared to Chronos-2, \sys is $8.2\times$ faster at embedding extraction and $1.5$--$2\times$ faster at retrieval. These advantages compound to $7.5\times$ over Chronos-2 and $45\times$ over TSFresh at $\mathrm{nprobe}=1$ (Figure~\ref{fig:search_scalability}).

\subsection{Ablation Study}
\label{subsec:ablation}
\label{subsec:design_validation}

\smartparagraph{The key design choices contribute independently.}
We ablate on Campus-Pkts-1 across all three granularities (Appendix Figure~\ref{fig:design_validation}).
\emph{Quantile tokenization} (replacing Chronos's uniform binning with quantile codebooks, identical architecture) consistently lowers MAPE and WD, with the largest gains at service level where burstiness is most extreme---confirming the codebook design of \S\ref{sec:representation} as the single most impactful component.
\emph{Twin-head architecture:} the shared-backbone twin-head outperforms the dual-model variant (independent IBG and BI encoders) across all granularities---MAPE drops from 0.044 to 0.008 and WD from 1.999 to 0.110 on service-level traces---confirming that joint encoding captures cross-stream dependencies that independent encoders miss.
The oracle decomposition (Figure~\ref{fig:design_validation}c) pinpoints burst-gap prediction as the harder sub-problem and the primary target for future improvements.
\emph{Threshold robustness:} sweeping $T_{\mathrm{act}}$ across percentiles (Figure~\ref{fig:thresh_sensitivity}) shows MAPE and WD stable over a broad range, degrading only at extreme thresholds where too few events remain for effective learning.

\vspace{-1em}

\section{Related Work}
\label{sec:related}

Classical forecasters---ARIMA~\cite{box1970time} and FARIMA~\cite{hosking1981farima}, the variant designed for long-memory processes like self-similar network traffic~\cite{leland2002self}---produce point predictions but no learned representations, limiting them to forecasting alone.
Deployed observability platforms (Grafana~ML~\cite{grafana_ml}, Elastic anomaly detection~\cite{elastic_anomaly}, Splunk~MLTK~\cite{splunk_mltk}) add binary anomaly labels but still lack clustering, characterization, or search~\cite{lakhina2005mining, mueller2025}.

Modern deep forecasters---DeepAR~\cite{deepar}, N-BEATS~\cite{n-beats}, Informer~\cite{zhou2021informer}, DLinear~\cite{zeng2023dlinear}, Autoformer~\cite{wu2021autoformer}, FEDformer~\cite{zhou2022fedformer}, TimesNet~\cite{wu2023timesnet}---encode mild-regime assumptions (dense, periodic) that wild-regime bursts violate.
Foundation models---Chronos~\cite{ansari2024chronos}, Lag-Llama~\cite{lagllama}, S2P2~\cite{chang2025s2p2}, and others~\cite{das2024timesfm, goswami2024moment, woo2024moirai, liu2024timer}---share three structural limitations: uniform tokenization erases tails, fixed 512-token contexts waste capacity on sparse traces, and instance normalization~\cite{liu2022non} attenuates heavy-tailed structure.
Chronos-2~\cite{ansari2025chronos2} addresses these with patch-based quantile regression, group attention for multivariate support, and 8{,}192-step context; Toto~\cite{toto} scales training to 1T+ observability points.
Temporal point processes (Hawkes~\cite{hawkes1971spectra}, Neural Hawkes~\cite{mei2017neuralhawkes}, THP~\cite{transformer_hawkes}, S2P2~\cite{chang2025s2p2}) separate timing from marks but model marks as categorical types, not continuous magnitudes.
All of the above optimize for prediction accuracy and treat the learned representation as a byproduct---a limitation we quantify in \S\ref{subsec:repr_quality} and \S\ref{subsec:forecasting}.

\sys takes a fundamentally different approach: rather than improving how a monolithic sequence is tokenized or scaled, eventization strips idle periods to compress context, and the twin-head encoder decomposes each trace into separate timing and magnitude streams---a structural decomposition that neither improved tokenization (Chronos-2) nor training scale (Toto) can replicate, because a monolithic sequence model must fit timing and magnitude with shared parameters regardless of how much data it trains on.

\cite{beltiukov2025demystifying} document ``cone collapse'' learned representations; \sys's soft cross-entropy addresses this problem by spreading probability mass across neighboring quantile bins, producing isotropic geometry (\S\ref{subsec:repr_quality}).
TSFresh~\cite{tsfresh} extracts ${\sim}$770 features per time series but its distances are dominated by features that take on large values, it does not support forecasting (\S\ref{subsec:forecasting}) or efficient search (\S\ref{subsec:search}), and its clusters are not isotropic (\S\ref{subsec:repr_quality}); \sys uses TSFresh as an attribution vocabulary via $I_c$ (\S\ref{sec:characterization}).

Decision trees~\cite{moshkovitz2020explainable, frost2020exkmc}, rule mining, and learned feature gates~\cite{svirsky2024interpretable} explain clusters through surrogate models but scale poorly to large cluster sizes $K$ because of the  hierarchical dependencies, the need for NP-hard optimization, or excessive retraining requirements~\cite{hu2024interpretable}.
\sys's $I_c$ avoids discrete structures entirely by combining continuous alignment and discrimination into a single scalar per cluster (\S\ref{sec:characterization}), scaling to $K\!=\!500$ without retraining or tree construction.
\section{Discussion}
\label{sec:discussion}
\label{sec:limitations}

\smartparagraph{Limitations.}
Three design choices limit the current scope.
(1)~\emph{Dataset-level quantization.} \sys applies $T_{\mathrm{act}}$ and quantile tokenization at dataset granularity rather than per entity (\S\ref{sec:representation}), explaining Chronos-2's stronger WD on the two mild-regime latency datasets. Developing per-entity $T_{\mathrm{act}}$ selection strategies that efficiently encode operator intent remains an open problem.
(2)~\emph{Burst-as-spike modeling.} \sys aggregates each burst into a single intensity value. Since forecasting, characterization, and search all operate on burst-level summaries, this suffices for the tasks we evaluate. In future work that concerns additional tasks, we plan to capture within-burst dynamics through richer kernels or learned burst profiles.
(3)~\emph{Statistical interpretability proxies.} The interpretability score $I_c$ (\S\ref{sec:characterization}) quantifies cluster describability via statistical proxies; calibrating it against human operator judgment is a natural next step but calls for a full-fledged user study. This would enable standard retrieval evaluations, such as recall@k against operator-validated notions of relevance.

\smartparagraph{Future directions.}
We evaluate three spatial aggregations---service (source~IP, destination~IP, and min of source and destination port), per-IP, and per-subnet---but many other choices exist (e.g., IP-pair, IP-subnet flow matrices). Each aggregation is expected to exhibit similar heavy-tailed, self-similar structure~\cite{leland2002self, willinger2002self, taqqu1997proof}, and the architecture should generalize; validating across additional aggregations is a natural first extension of this work.
Three further extensions would broaden overall scope.
(1)~\emph{Multivariate enrichment:} each entity currently contributes one time series (e.g., byte counts). Jointly encoding multiple modalities---bytes, packets, latency---for the same entity would capture richer per-entity behavior without changing the pipeline's univariate-first design.
(2)~\emph{Spatial composition:} \S\ref{sec:introduction} identified the tension between joint spatial-temporal modeling---whose cost grows with the network size---and independent per-entity learning. Composing \sys's embeddings with graph structure that propagates information along the network topology would close this gap.
(3)~\emph{Operational integration:} an LLM can query a time-series database to analyze a single entity, but similarity search across tens of thousands of entities or a taxonomy of hundreds of behavioral types requires compact, structured representations~\cite{feamster_why_2017, mestres2017knowledge}. \sys's per-entity embeddings and per-cluster feature descriptions provide the semantic bridge between raw telemetry data and an operator-recognizable vocabulary.
In a deployed setting, \sys can sit between telemetry collectors and operational tools, providing the structured summaries that dashboards, alert systems, and LLM-based agents need to move from ``something is anomalous'' to ``this is what it is and when it last happened.''

\section{Conclusion}

Compared to the strongest baselines and across nine production telemetry configurations, \sys achieves $1.3$--$116\times$ lower median forecasting error on wild-regime data, $1.0$--$7.5\times$ better match to the true burst distribution, $16\times$ higher median interpretability score, and $7.5\times$ faster end-to-end retrieval---while matching baselines on mild-regime benchmarks.
The core insight is that network telemetry alternates between idle stretches and operationally relevant bursts. \sys exploits this structure through an event-centric pipeline that converts a single intractable wild-regime problem into two simpler streams---each resembling the mild-regime data where existing models already handle well---by disentangling burst timing from burst magnitude.
By demonstrating that domain-aware input decomposition outperforms model scaling, \sys challenges the assumption that network telemetry demands ever-larger foundation models. As network operations move toward agent-driven automation, these agents need a perception layer that distills raw telemetry into faithful, compact representations of network state---\sys's event-centric embeddings are a step toward building that layer.

\section{Acknowledgement}
The authors thank Dr. Chris Misa (University of Oregon) for thoughtful feedback on our problem formulation and experiments. This work was supported in part by the National Science Foundation (CAREER Award No. 2443777 and CNS Award No. 2323229) and a research gift from Cisco. This research used resources of the National Energy Research Scientific Computing Center, a DOE Office of Science User Facility supported by the Office of Science of the U.S. Department of Energy under Contract No. DE-AC02-05CH11231 using NERSC award NERSC DDR-ERCAP0029768. We gratefully acknowledge access to NERSC GPU systems that enabled the large-scale training and evaluation in this work.

\bibliographystyle{plain}
\bibliography{nsdi27_netburst}

\appendix
\newpage
\section{Appendix}
\label{sec:appendix}

\subsection{Availability}
\label{sec:availability}
To support reproducibility, we will release an artifact package for \sys upon acceptance. Source code for the full experimental pipeline---telemetry preprocessing and eventization, model pretraining and fine-tuning, embedding extraction, clustering, and the FAISS-based retrieval pipeline---will be published on GitHub~\cite{github} under an open-source license. For retrieval, we will provide code for constructing the FAISS indices from the base embeddings used in the paper, as well as the query-time search pipeline used in our evaluation. The code will be present at \url{https://github.com/SNL-UCSB/NetBurst-essential}. We will also release the final model checkpoints and the preprocessed, anonymized datasets used for training and evaluation via Zenodo~\cite{zenodo}. Taken together, these artifacts are intended to enable reproduction of the paper's main results and to serve as a foundation for future research on event-centric representation learning for network telemetry---including extension to new data modalities, alternative encoder architectures, and operator-specific characterization pipelines.

\subsection{Baselines and Hyperparameters}
Table~\ref{tab:baselines} summarizes the baseline models and the hyperparameters used in our experiments. To keep the comparison consistent and faithful to prior work, we use the default hyperparameter settings recommended in the original papers or released implementations for each baseline unless otherwise noted. We apply the same training, validation, and evaluation protocol across all methods, and only make dataset-specific adjustments when required for compatibility with our forecasting setup (e.g., input length, prediction horizon, or optimization constraints). This design keeps the baselines close to their standard usage while ensuring a fair comparison with \sys.

\begin{table*}[ht]
  \caption{Baselines: architecture, input capacity, training mode, and key hyperparameters. All models use early stopping (patience 10) on validation loss. Context length and batch size are fixed across models for fair comparison; other settings follow each baseline's recommended configuration.}
  \label{tab:baselines}
  \centering
  \begin{small}
  \begin{tabular}{llcccccc}
    \textbf{Model} & \textbf{Category} & \textbf{Context} & \textbf{Mode} & \textbf{Hidden} & \textbf{Vocab.} & \textbf{Learn.\ Rate} & \textbf{Optim.} \\
    \midrule
    Chronos~\cite{ansari2024chronos}   & Foundation      & 512 tok  & Fine-tune & 512 & 4096 & $10^{-5}$ & Adam \\
    Chronos-2                            & Foundation      & 512 tok  & Fine-tune & 512 & 4096 & $10^{-5}$ & Adam \\
    Lag-Llama~\cite{lagllama} & Foundation   & 512 tok  & Fine-tune & 144 & ---  & $10^{-5}$ & Adam \\
    Toto~\cite{toto}                   & Foundation      & 512 tok  & Zero-shot & 768  & ---  & ---       & ---  \\
    \midrule
    DeepAR~\cite{deepar}               & Deep forecaster & $2H$--$7H$ & Scratch & 64  & ---  & $10^{-3}$ & Adam \\
    N-BEATS~\cite{n-beats}             & Deep forecaster & $2H$--$7H$ & Scratch & 512 & ---  & $10^{-3}$ & Adam \\
    \midrule
    THP~\cite{transformer_hawkes}      & Point process   & Full seq.  & Scratch & 256  & ---  & $1e^{-3}$       & Adam  \\
    S2P2                                & Point process      & Full seq  & Scratch      & 64  & ---  & $1e^{-3}$       & Adam  \\
    \midrule
    TSFresh~\cite{tsfresh}             & Stat.\ features & 770 feat.  & ---     & ---  & ---  & ---       & ---  \\
    \midrule
    \sys                                & Event-centric   & 512 tok  & Scratch & 512 & 4096 & $10^{-3}$ & Adam \\
  \end{tabular}
  \end{small}
\end{table*}

\subsection{Per-Entity Forecasting Distributions}

\begin{figure*}[!ht]
    \centering

    \begin{subcaptionbox}{Campus-Pkts-1 IP}[0.32\textwidth]
        {\includegraphics[width=\linewidth]{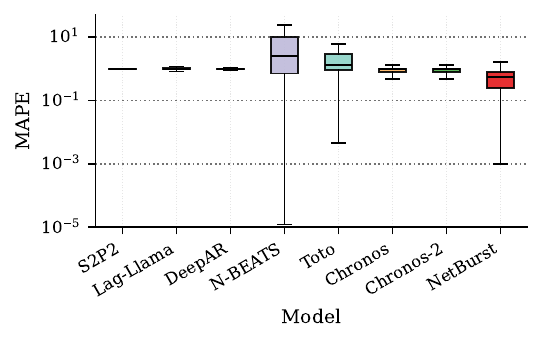}}
    \end{subcaptionbox}
    \hfill
    \begin{subcaptionbox}{Campus-Pkts-1 Service}[0.32\textwidth]
        {\includegraphics[width=\linewidth]{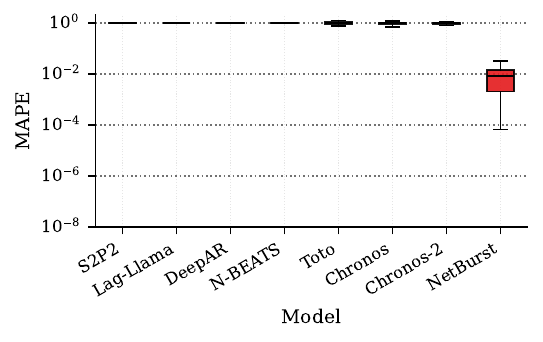}}
    \end{subcaptionbox}
    \hfill
    \begin{subcaptionbox}{Campus-Pkts-1 Subnet}[0.32\textwidth]
        {\includegraphics[width=\linewidth]{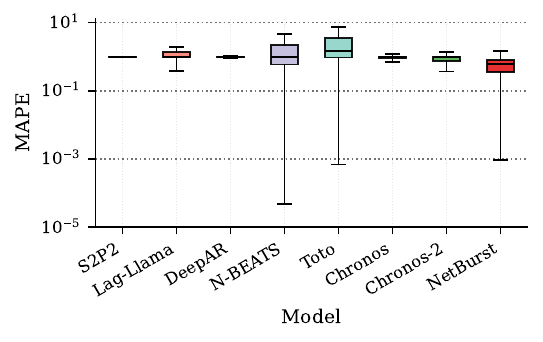}}
    \end{subcaptionbox}

    \par\medskip

    \begin{subcaptionbox}{MAWI IP}[0.32\textwidth]
        {\includegraphics[width=\linewidth]{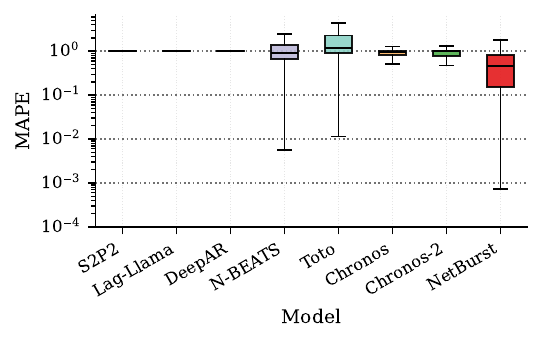}}
    \end{subcaptionbox}
    \hfill
    \begin{subcaptionbox}{MAWI Service}[0.32\textwidth]
        {\includegraphics[width=\linewidth]{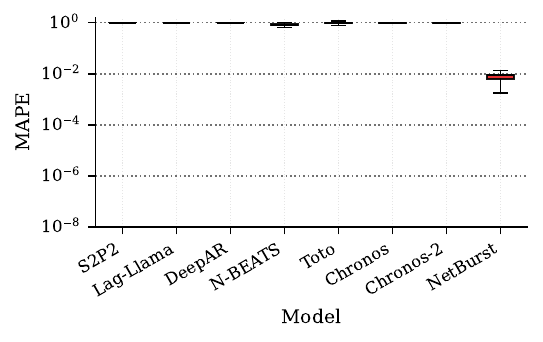}}
    \end{subcaptionbox}
    \hfill
    \begin{subcaptionbox}{MAWI Subnet}[0.32\textwidth]
        {\includegraphics[width=\linewidth]{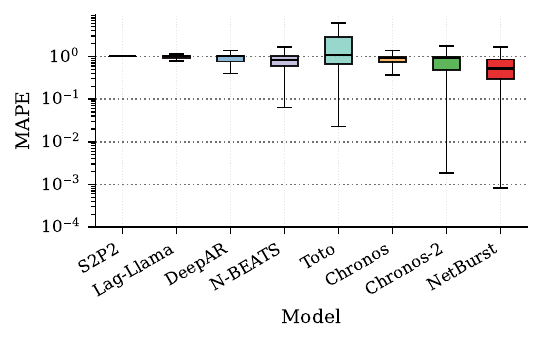}}
    \end{subcaptionbox}

    \par\medskip

    \begin{subcaptionbox}{RnE NetFlow}[0.32\textwidth]
        {\includegraphics[width=\linewidth]{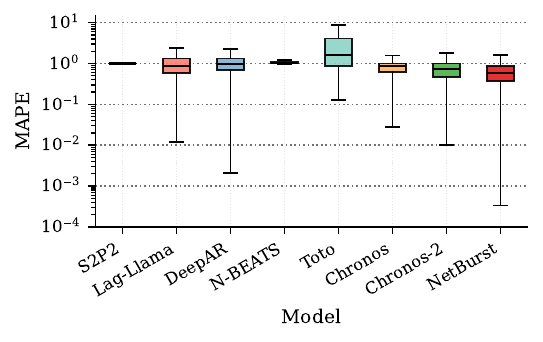}}
    \end{subcaptionbox}
    \hfill
    \begin{subcaptionbox}{RnE Latency}[0.32\textwidth]
        {\includegraphics[width=\linewidth]{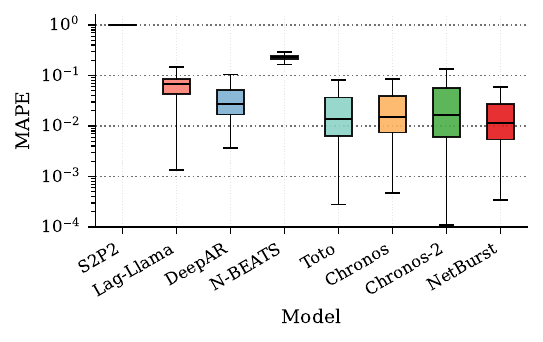}}
    \end{subcaptionbox}
    \hfill
    \begin{subcaptionbox}{Campus Latency}[0.32\textwidth]
        {\includegraphics[width=\linewidth]{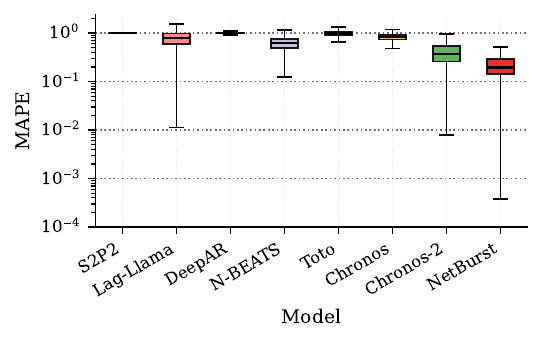}}
    \end{subcaptionbox}

    \caption{Per-entity MAPE distributions across 9 datasets. Each box plot shows the distribution of per-entity MAPE values for a single model. Lower is better. THP is omitted because its errors are an order of magnitude larger and compress the remaining models into an unreadable range. \sys (rightmost) achieves the lowest median MAPE on all packet-based datasets and remains competitive on latency datasets.}
    \label{fig:appendix_mape_box_3x3}
\end{figure*}

\begin{figure*}[ht!]
    \centering

    \begin{subcaptionbox}{Campus-Pkts-1 IP}[0.32\textwidth]
        {\includegraphics[width=\linewidth]{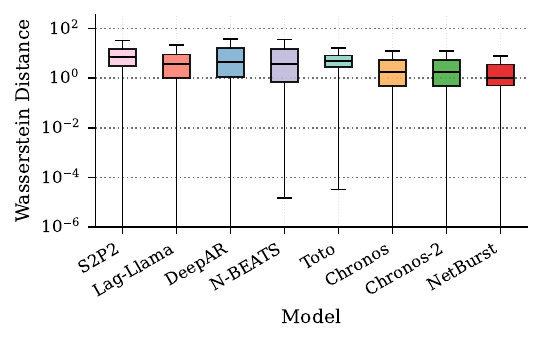}}
    \end{subcaptionbox}
    \hfill
    \begin{subcaptionbox}{Campus-Pkts-1 Service}[0.32\textwidth]
        {\includegraphics[width=\linewidth]{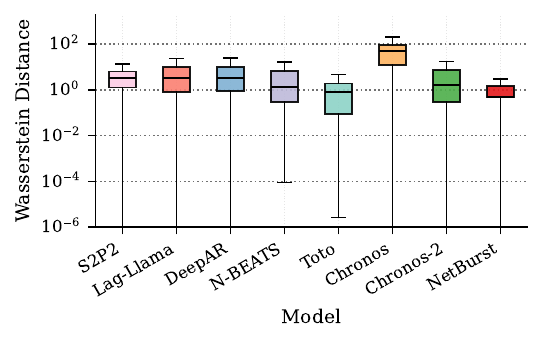}}
    \end{subcaptionbox}
    \hfill
    \begin{subcaptionbox}{Campus-Pkts-1 Subnet}[0.32\textwidth]
        {\includegraphics[width=\linewidth]{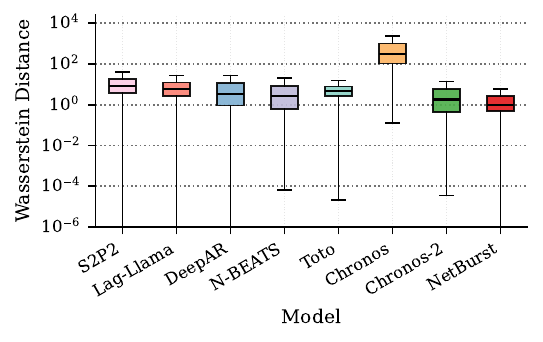}}
    \end{subcaptionbox}

    \par\medskip

    \begin{subcaptionbox}{MAWI IP}[0.32\textwidth]
        {\includegraphics[width=\linewidth]{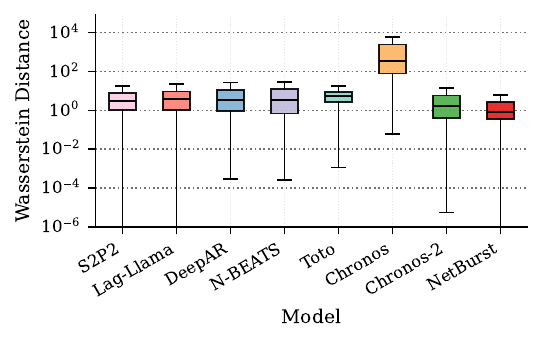}}
    \end{subcaptionbox}
    \hfill
    \begin{subcaptionbox}{MAWI Service}[0.32\textwidth]
        {\includegraphics[width=\linewidth]{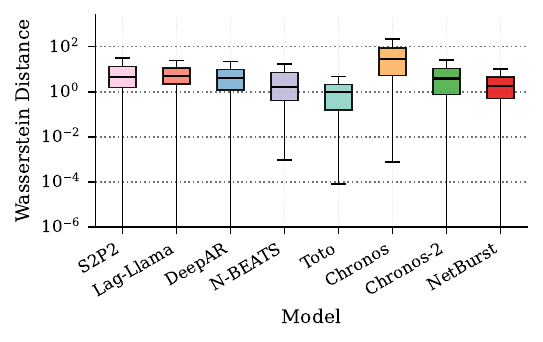}}
    \end{subcaptionbox}
    \hfill
    \begin{subcaptionbox}{MAWI Subnet}[0.32\textwidth]
        {\includegraphics[width=\linewidth]{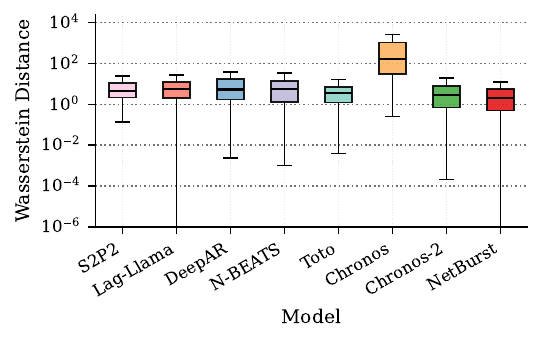}}
    \end{subcaptionbox}

    \par\medskip

    \begin{subcaptionbox}{RnE NetFlow}[0.32\textwidth]
        {\includegraphics[width=\linewidth]{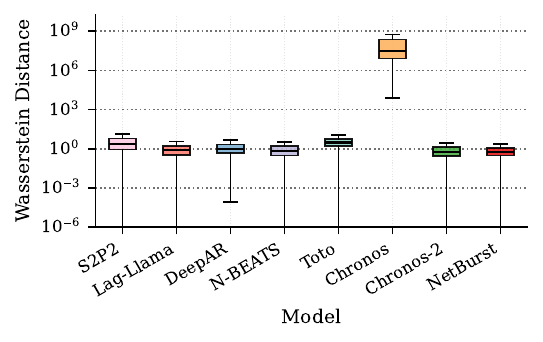}}
    \end{subcaptionbox}
    \hfill
    \begin{subcaptionbox}{RnE Latency}[0.32\textwidth]
        {\includegraphics[width=\linewidth]{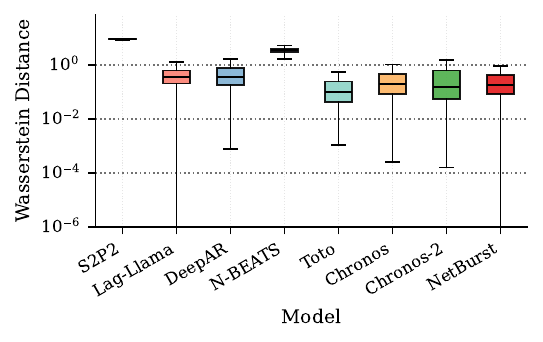}}
    \end{subcaptionbox}
    \hfill
    \begin{subcaptionbox}{Campus Latency}[0.32\textwidth]
        {\includegraphics[width=\linewidth]{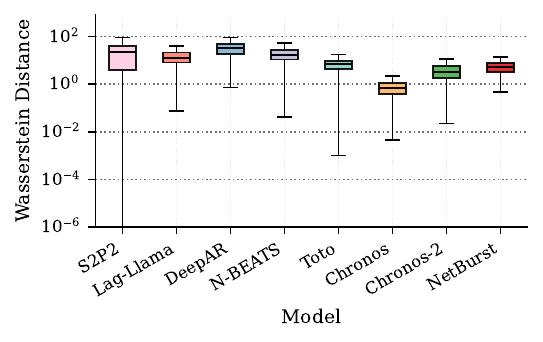}}
    \end{subcaptionbox}

    \caption{Per-entity Wasserstein distance (WD) distributions across 9 datasets. Each box plot shows the distribution of per-entity WD values for a single model. Lower WD indicates better distributional fidelity between predicted and observed traffic. THP is omitted (see Figure~\ref{fig:appendix_mape_box_3x3} caption). \sys achieves the tightest WD distributions on packet-based datasets, confirming that its forecasts preserve the shape of the traffic distribution, not just the mean.}
    \label{fig:appendix_wd_box_3x3}
\end{figure*}

\autoref{fig:appendix_mape_box_3x3}-\ref{fig:appendix_wd_box_3x3} show box plots of the success metrics MAPE and WD for all evaluated models across all datasets. We omit the Transformer Hawkes Process (THP) from this visualization because its errors are substantially larger than those of the remaining models and including it would %
obscure the comparisons of interest. %

\label{sec:appendix_forecasting_distributions}

\newpage

\newpage
\subsection{Ablation of Architectural Choices}
\label{app:ablation_architecture}

\begin{figure*}[h]
    \centering
    \begin{subcaptionbox}{MAPE under different architectural choices. \label{fig:ablation_mape}}[0.32\textwidth]
        {\includegraphics[width=\linewidth]{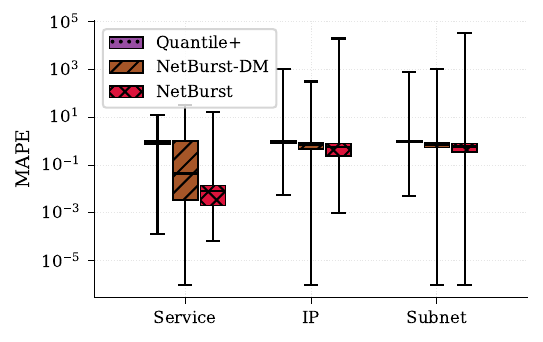}}
    \end{subcaptionbox}
    \begin{subcaptionbox}{WD under different architectural choices. \label{fig:ablation_wd}}[0.32\textwidth]
        {\includegraphics[width=\linewidth]{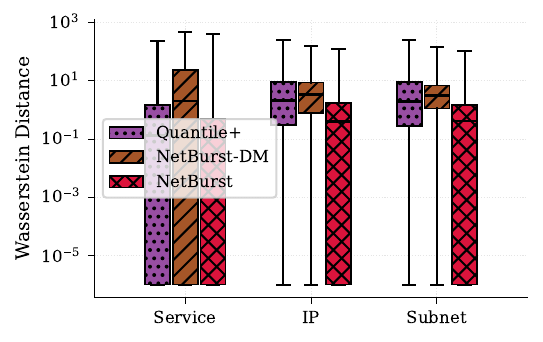}}
    \end{subcaptionbox}
    \begin{subcaptionbox}{WD under different oracles. \label{fig:ablation_oracle}}[0.32\textwidth]
        {\includegraphics[width=\linewidth]{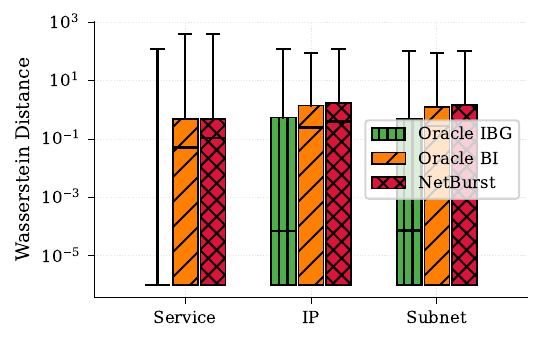}}
    \end{subcaptionbox}
    \caption{Ablation on Campus-Pkts-1 at IP, service, and subnet granularities (lower is better in all panels). (a,b)~Three architectural variants are compared: \textbf{Quantile+} applies quantile tokenization on the Chronos backbone without eventization; \textbf{\sys-DM} adds eventization but keeps IBG and BI as separate models; \textbf{\sys} fuses both streams in a shared-backbone twin-head encoder. Each step yields a measurable improvement---at service level, MAPE drops from 0.044 (Quantile+) to 0.008 (\sys) and WD from 1.999 to 0.110---confirming that eventization and cross-stream fusion each contribute independently (\S\ref{subsec:ablation}). (c)~Oracle decomposition on WD: \emph{BI oracle} feeds ground-truth burst intensity with predicted IBG; \emph{IBG oracle} feeds ground-truth inter-burst gaps with predicted BI. The larger gap for the BI oracle shows that burst-gap prediction is the harder sub-problem and the primary target for future improvement. MAPE oracles are omitted because Oracle IBG leaves BI unchanged (same MAPE as \sys) and Oracle BI uses ground-truth BI by definition.}
    \label{fig:design_validation}
\end{figure*}

To isolate the contribution of each component of our architecture, we compare \sys against two progressively stronger ablations and show the results in Figure~\ref{fig:design_validation}. First, we remove eventization entirely and retain only quantile-based discretization. We denote this variant as \textbf{Quantile+}. This ablation tests whether quantile binning alone is sufficient to model the heavy-tailed behavior of network telemetry, and the results are shown in see Figure~\ref{fig:design_validation}(a).

Next, we enable both eventization and quantile-based binning, but keep the inter-burst gap (IBG) and burst intensity (BI) streams as separate models rather than combining them within a shared backbone. We denote this variant as \textbf{\sys-DM} (\sys Dual Model). This ablation isolates the value of eventization while omitting cross-stream fusion; see Figure~\ref{fig:design_validation}(b).

The full \sys architecture goes one step further by fusing the BI and IBG representations within a single shared backbone---a single encoder whose layers process both streams jointly before splitting into separate prediction heads at the output. This design allows the model to jointly capture the interaction between burst timing and burst magnitude, rather than learning them independently. As reported in \S\ref{subsec:ablation}, this progression yields consistent gains: at service level, MAPE drops from 0.044 (Quantile+) to 0.008 (\sys) and WD from 1.999 to 0.110, confirming that eventization and cross-stream fusion each contribute independently.

We also use an oracle decomposition analysis to separate the difficulty of the two prediction tasks. In this analysis, one component is replaced with its ground-truth oracle value while the model predicts the other---for example, replacing predicted IBG with observed IBG and measuring BI prediction error in isolation---allowing us to measure how errors in each stream propagate to the final reconstruction. %
Figure~\ref{fig:design_validation}(c) shows that burst-gap prediction is the harder sub-problem and remains the primary target for future improvements. This result supports the design of \sys: while eventization and fusion both improve performance, the dominant remaining challenge lies in accurately modeling burst timing rather than burst magnitude alone.

\newpage

\newpage
\subsection{Threshold sensitivity}

\begin{figure}[H]
    \centering
    \begin{subcaptionbox}{MAPE vs.\ threshold. \label{fig:thresh_ip_mape}}[0.48\linewidth]
        {\includegraphics[width=\linewidth]{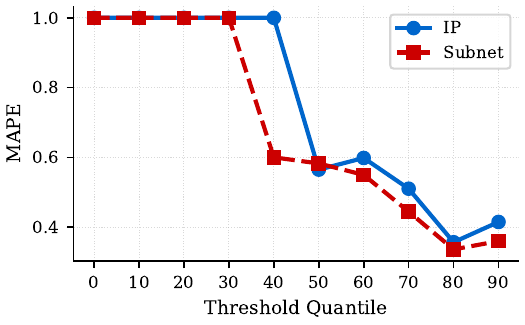}}
    \end{subcaptionbox}\hfill
    \begin{subcaptionbox}{WD vs.\ threshold. \label{fig:thresh_ip_wd}}[0.48\linewidth]
        {\includegraphics[width=\linewidth]{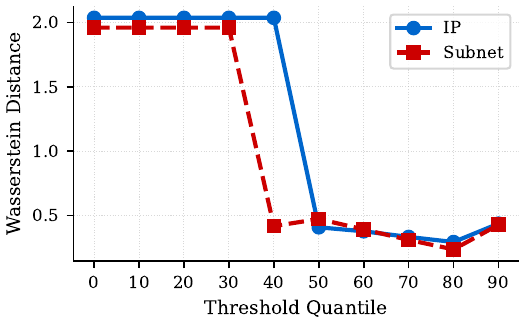}}
    \end{subcaptionbox}
    \caption{Threshold sensitivity on Campus-Pkts-1: MAPE and WD as a function of the activity threshold $T_{\mathrm{act}}$ percentile for IP- and subnet-level aggregations. Both metrics are flat across low percentiles (where many thresholds map to zero), improve as the threshold suppresses minor fluctuations, then degrade sharply at the $90^{\mathrm{th}}$ percentile where too few events survive for effective learning. The broad plateau comprising the 50th–80th percentile range shows that \sys is robust to $T_{\mathrm{act}}$ selection (\S\ref{subsec:ablation}).}
    \label{fig:thresh_sensitivity}
\end{figure}

Figure \ref{fig:thresh_sensitivity} shows the sensitivity of our approach and model architecture to the choice of the $T_{act}$ Threshold. We observe that both MAPE and WD are initially flat because the lower-percentile thresholds are unchanged: the \(0^\text{th}\)–\(40^\text{th}\) percentile thresholds for Campus-Pkts-1 IP are all 0, and the \(0^\text{th}\)–\(30^\text{th}\) percentile thresholds for Campus-Pkts-1 Subnet are likewise 0. As the threshold increases beyond this range, both metrics initially improve, suggesting that \sys benefits from suppressing minor fluctuations and concentrating on stronger, more meaningful signals. However, further increasing the threshold reduces the number of events available for learning and prediction. At the \(90^{\mathrm{th}}\) percentile, both MAPE and WD increase, indicating that overly aggressive thresholding removes useful predictive signal.

\newpage
\subsection{Chronos-2 Cluster Interpretability Details}

\begin{figure*}[p!]
    \centering
    \begin{subfigure}[t]{0.2\textwidth}
        \centering
        \includegraphics[width=\linewidth]{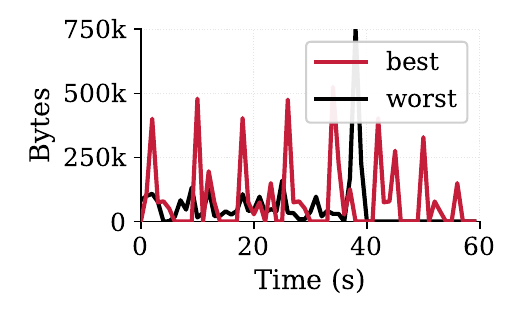}
        \caption{Rank 1}
        \label{fig:c2_taxonomy_rank1}
    \end{subfigure}%
    \hfill
    \begin{subfigure}[t]{0.2\textwidth}
        \centering
        \includegraphics[width=\linewidth]{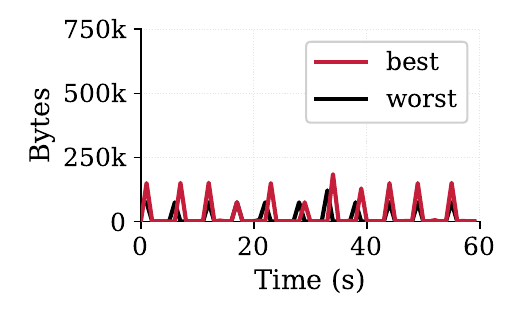}
        \caption{Rank 2}
        \label{fig:c2_taxonomy_rank2}
    \end{subfigure}%
    \hfill
    \begin{subfigure}[t]{0.2\textwidth}
        \centering
        \includegraphics[width=\linewidth]{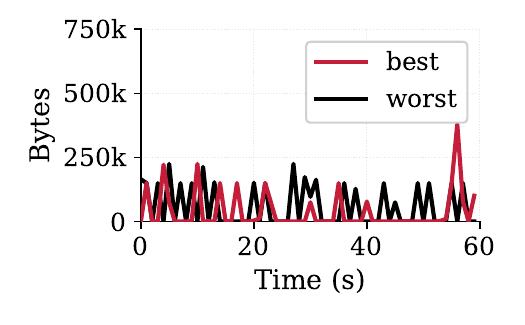}
        \caption{Rank 250}
        \label{fig:c2_taxonomy_rank250}
    \end{subfigure}%
    \hfill
    \begin{subfigure}[t]{0.2\textwidth}
        \centering
        \includegraphics[width=\linewidth]{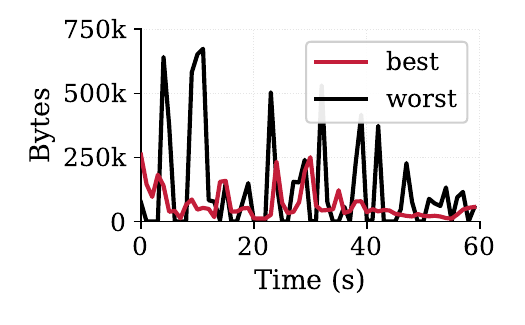}
        \caption{Rank 499}
        \label{fig:c2_taxonomy_rank499}
    \end{subfigure}%
    \hfill
    \begin{subfigure}[t]{0.2\textwidth}
        \centering
        \includegraphics[width=\linewidth]{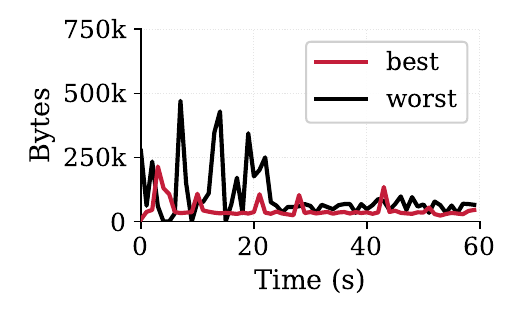}
        \caption{Rank 500}
        \label{fig:c2_taxonomy_rank500a}
    \end{subfigure}%
    \caption{Chronos-2 cluster overlays for the five clusters in Table~\ref{tab:selected_cluster_taxonomy_chronos2}. For each cluster, two members are randomly sampled and overlaid: the member whose feature values best match the cluster average (\textcolor{red}{red}) and the member whose values diverge most (\textbf{black}). At rank~1, the two traces look fundamentally different---one bursty, one flat---yet Chronos-2 placed them in the same cluster, illustrating why its $I_c$ scores are low. Compare with \sys's overlays in Figure~\ref{fig:cluster_examples}, where high-$I_c$ clusters produce visually coherent pairs (\S\ref{subsec:ablation}).}
    \label{fig:cluster_examples_chronos2}
\end{figure*}

To evaluate how describable the clusters obtained by Chronos-2 are, we apply the same methodology we used in \S\ref{sec:characterization} to examine \sys's clusters.  Table~\ref{tab:selected_cluster_taxonomy_chronos2} lists the top 2, median, and bottom 2 Chronos-2 clusters by interpretability score $I_c$. Figure~\ref{fig:cluster_examples_chronos2} shows time series that best and worst align with each cluster's description.

\begin{table*}[p!]
\vspace{-1em}
\caption{Natural-language descriptions of five selected Chronos-2 clusters from the $I_c$ distribution ($K\!=\!500$): the top-2 (highest $I_c$), one median cluster, and the bottom-2 (lowest $I_c$). Per-cluster top-5 features ranked by $\mathrm{CKA}\cdot\tilde{d}$ with cluster means. Feature notation is defined below.$^{\dagger}$}
  \begin{minipage}{\textwidth}
  \vspace{1pt}
  \footnotesize $^{\dagger}$Unless prefixed, each feature is \texttt{change\_quantiles}: variance of step-to-step changes restricted to values whose percentile rank falls in $[q_a,q_b]$. ``$\pm$'' = signed; ``$|\cdot|$'' = absolute. Other prefixes: \texttt{count\_above/below\_mean} = samples above/below the mean; \texttt{peaks3} = peaks with support $n\!=\!3$; \texttt{CV} = coefficient of variation; \texttt{IMQ}$_{.9}$ = index mass quantile at $q\!=\!.9$; \texttt{LZ} = Lempel--Ziv complexity; \texttt{PE} = permutation entropy; \texttt{c3} = third-order nonlinear dependence statistic (lag~3).
  \end{minipage}
\label{tab:selected_cluster_taxonomy_chronos2}
\centering
\begin{small}
\setlength{\tabcolsep}{4pt}
\renewcommand{\arraystretch}{1.0}
\begin{tabular}{@{}c>{\raggedright\arraybackslash}p{1.8cm}>{\raggedright\arraybackslash}p{5.0cm}>{\raggedright\arraybackslash}p{9.0cm}@{}}
\textbf{Rank} & \textbf{Score ($I_c$)} & \textbf{Top-5 features (cluster mean)} & \textbf{Natural language description} \\
\midrule
1 & 0.0575
& $|\cdot|[.6,.8]{:}\,1030.9$;\, $\pm[.6,.8]{:}\,2221.9$;\, \texttt{c3}${:}\,3.60{\times}10^8$;\, $|\cdot|[.0,.8]{:}\,3755.2$;\, $|\cdot|[.2,.8]{:}\,3755.2$
& High-amplitude burst transitions with strong nonlinear dependence between consecutive values, concentrated in the upper range of observed traffic levels. These features capture upper-range volatility and nonlinear temporal coupling. \\
2 & 0.00319
& $|\cdot|[.0,.6]{:}\,261.9$;\, $|\cdot|[.2,.6]{:}\,261.9$;\, $\pm[.0,.6]{:}\,360.3$;\, $\pm[.2,.6]{:}\,360.3$;\, $|\cdot|[.0,.4]{:}\,6.55$
& Moderate volatility concentrated in the mid-range of observed values, with negligible transitions among the lowest values. These features capture where step-to-step changes concentrate within the value distribution. \\
250 & 0.00320
& $\pm[.0,.6]{:}\,118.3$;\, $\pm[.2,.6]{:}\,118.3$;\, $\pm[.4,.6]{:}\,118.3$;\, $|\cdot|[.0,.6]{:}\,89.4$;\, $|\cdot|[.2,.6]{:}\,89.4$
& Moderate volatility originating from a single narrow band in the mid-upper range of observed values. These features capture where step-to-step transitions concentrate within the value distribution. \\
499 & $3.22{\times}10^{-5}$
& $|\cdot|[.2,.6]{:}\,4.73{\times}10^5$;\, $\pm[.2,.6]{:}\,8.64{\times}10^5$;\, \texttt{count\_above\_mean}${:}\,135.6$;\, \texttt{count\_below\_mean}${:}\,464.4$;\, \texttt{peaks3}${:}\,65.7$
& Right-skewed traffic where most values fall below the mean, with frequent local peaks and large step-to-step transitions in the mid-range. These features capture distribution asymmetry, local peak density, and mid-range variability. \\
500 & $3.87{\times}10^{-5}$
& \texttt{CV}${:}\,1.86$;\, \texttt{IMQ}$_{.9}{:}\,0.857$;\, \texttt{LZ}${:}\,0.363$;\, \texttt{PE}$_{d=3}{:}\,1.72$;\, \texttt{PE}$_{d=7}{:}\,5.94$
& Highly variable traffic where most of the total volume arrives in the first portion of the observation window, with moderately complex temporal ordering. These features capture relative variability, mass concentration in time, and sequence complexity. \\
\end{tabular}
\end{small}
\end{table*}

\subsection{LLM Usage}
\label{sec:llm_prompt}
Portions of this work benefited from the use of large language models (LLMs). LLM-based code editors were employed to expedite prototyping and experimentation, and interactive chat-based interfaces assisted in refining the manuscript text. All substantive ideas, methodological contributions, and experimental designs originate from the authors.

\smartparagraph{Cluster description generation.}
The natural-language descriptions in Tables~\ref{tab:cluster_taxonomy} and~\ref{tab:selected_cluster_taxonomy_chronos2} are generated by prompting Claude Opus~4.6~\cite{anthropic2025claude} (Anthropic) with the following fixed prompt, applied identically to every cluster:

\begin{figure}[H]
\centering
\fbox{\begin{minipage}{0.92\columnwidth}
\small
\texttt{You are describing a cluster of network traffic time series for a network operator. You receive the top-5 statistical features that define this cluster, each with its average value across cluster members.}

\medskip
\texttt{Produce exactly TWO sentences:}\\
\texttt{- Sentence 1: Describe the traffic behavior this cluster represents in plain language a network operator would understand. Do not use feature names.}\\
\texttt{- Sentence 2: State which aspects of the traffic (e.g., burstiness, persistence, variability, tail behavior, periodicity) the features capture.}

\medskip
\texttt{Do not reference percentile values, corridor ranges, or feature implementation details. Do not hedge or speculate. Be concrete and direct.}
\end{minipage}}
\end{figure}

\noindent The input to each call consists solely of the five feature names and their cluster-member averages from the corresponding table row. To mitigate the inherent variability of LLM outputs, we regenerate each description 10 times with the same prompt and features, identify the phrases and behavioral characterizations that remain stable across iterations, and select the description whose wording most closely matches the consensus. This procedure ensures that the reported descriptions reflect the features' stable semantic content rather than a single stochastic sample.

\end{document}